\documentclass[twocolumn]{aastex7}
\usepackage{enumitem}
\usepackage{amsmath}

\newcommand{\myterm}[1]{#1}

\begin{document}

\title{Magnetic Fields and Asymmetric Accretion in the Class 0 Protostellar System L1527 IRS}

\author[0009-0004-5250-7302]{Hanju Nam}
\affiliation{Department of Science Education (Earth Science Major), Seoul National University, 1 Gwanak-ro, Gwanak-gu, Seoul 08826, Republic of Korea}
\email[]{skagkswn@snu.ac.kr}

\author[0000-0003-4022-4132]{Woojin Kwon}
\affiliation{Department of Earth Science Education, Seoul National University, 1 Gwanak-ro, Gwanak-gu, Seoul 08826, Republic of Korea}
\affiliation{Department of Science Education (Earth Science Major), Seoul National University, 1 Gwanak-ro, Gwanak-gu, Seoul 08826, Republic of Korea}
\affiliation{SNU Astronomy Research Center, Seoul National University, 1 Gwanak-ro, Gwanak-gu, Seoul 08826, Republic of Korea}
\affiliation{The Center for Educational Research, Seoul National University, 1 Gwanak-ro, Gwanak-gu, Seoul 08826, Republic of Korea}
\email[show]{wkwon@snu.ac.kr}

\author[0009-0004-9279-780X]{Youngwoo Choi}
\affiliation{Department of Physics and Astronomy, Seoul National University, Seoul 08826, Republic of Korea}
\email[]{cyw3614@snu.ac.kr}

\author[0000-0003-3017-4418]{Ian Stephens}
\affiliation{Department of Earth, Environment, and Physics, Worcester State University, Worcester, MA 01602, USA}
\email[]{iwstephens@gmail.com}

\author[0000-0002-4540-6587]{Leslie Looney}
\affiliation{Department of Astronomy, University of Illinois, 1002 West Green St, Urbana, IL 61801, USA}
\email[]{lwl@illinois.edu}

\correspondingauthor{Woojin Kwon}

\begin{abstract}
Magnetic fields play a crucial role in regulating the collapse of dense molecular clouds, shaping the early stages of protostellar evolution. We present polarization observations in submillimeter wavelengths with the SCUBA-2/POL-2 instrument on James Clerk Maxwell Telescope toward the Class 0 protostellar system L1527 IRS (IRAS 04368+2557). The magnetic field morphology varies across the core of L1527 IRS. Magnetic fields in the eastern region are perpendicular to the outflow axis, while those in the western region show a pinched morphology, overall aligned with the outflow cavity. These distinct bipolar outflow regions also have a clear color difference in near infrared observed by James Webb Space Telescope. In addition, the spectral index derived from 450 $\mu$m and 850 $\mu$m observations reveals that the northwestern region of the protostellar system is colder and denser compared to the southeastern region. Furthermore, in the large-scale structures and magnetic fields revealed by Herschel and Planck, we find a $\sim$0.1 pc scale filamentary structure parallel to the large-scale magnetic fields in the eastern region and a relatively isotropic mass distribution in the western region. Based on these results, we propose an accretion scenario of the protostellar system in which an asymmetric mass distribution causes the distinct features observed in near-infrared and submillimeter wavelengths.
\end{abstract}



\section{Introduction} \label{sec:intro}
Star formation begins when dense molecular clouds undergo gravitational collapse. However, not all clouds become stellar nurseries; only those in which gravity dominates over other regulating forces can initiate the process. Among the mechanisms proposed to regulate star formation, magnetic fields \citep[e.g.,][]{Shu, Mouschovias} and turbulence \citep[e.g.,][]{MacLow} are thought to play important roles in the star formation process. Magnetic fields play crucial roles in star formation, such as providing magnetic support against gravitational collapse and reducing angular momentum through magnetic braking. In addition, magnetic fields are closely related to other young stellar object (YSO) structures such as protostellar jets and outflows. Therefore, studying magnetic fields is crucial for understanding the overall star formation process.

L1527 IRS (hereafter L1527) is one of the most well-known and closest Class 0 protostars. It is located in the L1527 dark cloud of the Taurus star-forming region (R.A. = 4h 39min 53.9s, Dec. = $26^{\circ}$ $03^{\prime}$ 09\farcs8 in ICRS). \citet{Luhman} measured the distance to L1527 to be 139--141 pc based on Gaia Data Release 2 (DR2), consistent with the estimate of \citet{Zucker}. In this paper, we adopt 140 pc as a distance to the L1527. \cite{Tobin2012} showed that its envelope contains approximately 1 $M_{\odot}$ within a radius of roughly 10000 au.  The protostar has a mass of $\sim0.45\,M_{\odot}$, dynamically inferred from a nearly edge-on Keplerian rotating disk that is almost perfectly aligned in the north–south direction \citep[][]{Aso_2017, eDisk_L1527}. A bipolar outflow is also identified in both submillimeter and infrared observations. (Sub)millimeter CO observations show that L1527 has a bipolar outflow with an orientation nearly on the plane of the sky \citep{Hogerheijde, eDisk_L1527}, which is perpendicular to the Keplerian disk. Additionally, Spitzer Space Telescope IRAC \citep{Tobin2008} and James Webb Space Telescope (JWST) NIRCam observations \citep{L1527_JWST} clearly show a bipolar outflow cavity with an opening angle of $45^{\circ}$.

Magnetic fields of L1527 have been investigated by many previous studies on various scales. For interferometric observations, \cite{L1527_CARMA} presented the magnetic field of the central region using the CARMA 1.3 mm dust polarization observations with a resolution of 3$^{\prime\prime}$ ($\sim$420 au). \cite{L1527_Harris} investigated the dust polarization in L1527 at the disk-scale (0\farcs1 resolution, $\sim$14 au) using ALMA 875~$\mu$m observations and reported that the polarized signals on these scales are dominated by dust self-scattering rather than magnetically aligned dust grains. For single-dish observations, \cite{Davidson_2011} presented 350 $\mu$m observations using Caltech Submillimeter Observatory SHARP polarimeter with an effective beam size of $10^{\prime\prime}$ ($\sim$1400 au). However, this result only covered a small region of L1527 with a limited sensitivity \citep[e.g., Figure 2 in ][]{Davidson_2015}. \cite{Matthews_SCUPOL} presented another intermediate scale magnetic fields using James Clerk Maxwell Telescope (JCMT) 850 $\mu$m SCUBA observations with the SCUPOL polarimeter, with an effective beam size of $20^{\prime\prime}$ ($\sim$2800 au). However, the observation data sets were also obtained with a limited sensitivity and shows almost noisy pattern (see App.~\ref{app:comparison}). Therefore, high-sensitivity observations of intermediate-scale magnetic fields are still lacking for L1527. Instead of the SCUBA/SCUPOL, the next generation JCMT instruments, SCUBA-2 and POL-2, have provided more advanced performances successfully \citep[e.g.,][]{BISTRO}. In this paper, we present the intermediate-scale magnetic fields of L1527 using the JCMT SCUBA-2/POL-2 observation data with an effective beam size of 14\farcs6 ($\sim2000$ au). 

We describe the observations in Section~\ref{sec:obs}. Section~\ref{sec:results} presents the results of magnetic field morphology, intensity, and spectral index distributions. In Section~\ref{sec:discussion}, we present analyses of magnetic fields together with supplementary large-scale observational data from Herschel and Planck. Based on the results, we propose a star formation scenario of L1527. Conclusions are given in Section~\ref{sec:conclusion}. \\

\section{Observations}\label{sec:obs}
\subsection{JCMT SCUBA-2/POL-2}\label{subsec:obs_POL2}
L1527 was observed with the JCMT SCUBA-2/POL-2 instruments, using the POL-2-DAISY observing mode. The SCUBA-2/POL-2 provides linear polarization data with Stokes parameters $I$, $Q$, and $U$ at 450 $\mu$m and 850 $\mu$m simultaneously. The effective beam sizes of the multi-bands are 9\farcs8 ($\sim$1400 au) and 14\farcs6 ($\sim$2000 au), respectively. A total of 25 data sets were obtained in 2017 and 2023, with 8 sets taken in 2017 and 17 in 2023 (PI: Woojin Kwon, Proposal IDs: M17AP073, M23BP052). We followed common polarization data reduction process by using $pol2map$ in SMURF \citep{SMURF} of the Starlink software \citep{Starlink} supported by the East Asian Observatory. The final Stokes maps have a pixel size of $4^{\prime\prime} \times 4^{\prime\prime}$, and their unit is mJy~beam$^{-1}$. The rms noise levels in Stokes $I$ at 450~$\mu$m and 850~$\mu$m are 38.2~mJy~beam$^{-1}$ and 3.7~mJy~beam$^{-1}$, respectively. When deriving magnetic field orientations, we used the Stokes $I$, $Q$, and $U$ maps that were binned to a size of $12^{\prime\prime}\times12^{\prime\prime}$ at the last $pol2map$ stage to obtain a higher S/N. Polarization intensities were calculated as following equation to correct for the positive bias: $PI=\sqrt{Q^2+U^2 - \sigma^2_{PI}}$, where $\sigma^2_{PI}=(Q^2\sigma^2_{Q}+U^2\sigma^2_{U})/(Q^2+U^2)$. Polarization angles and fractions were computed using the equations: $\chi=0.5\,\text{tan}^{-1}(U/Q)$ and $p =PI/I$. The resulting rms noise levels in Stokes $I$ and polarization intensity are 7.8~mJy~beam$^{-1}$ and 8.2~mJy~beam$^{-1}$ at 450~$\mu$m, and 0.75~mJy~beam$^{-1}$ and 0.74~mJy~beam$^{-1}$ at 850~$\mu$m, respectively. We only used 850 $\mu$m polarization data to infer the magnetic field orientations, as it offers better sensitivity and reliable polarization calibration than the 450 $\mu$m observations. However, we also used the 450 $\mu$m Stokes $I$ data together for intensity and spectral index analyses in Section~\ref{sec:results}.\\

\subsection{JCMT HARP}\label{subsec:obs_HARP}
We observed $\mathrm{C}^{18}\mathrm{O}$ ($J=3\text{--}2$, $\nu_{\mathrm{rest}}=329.33056$~GHz) with the JCMT HARP/ACSIS instruments to investigate the envelope kinematics of L1527 (PI: Woojin Kwon, Proposal ID: M21BP074). The observations were carried out on 2021 November 12 in scan observing mode with a spectral resolution of 0.056 km~s$^{-1}$. The data were reduced using the ORAC-DR pipeline and the Kernel Application Package \citep[KAPPA;][]{kappa} in Starlink. The effective beam size is 14\farcs6 with Nyquist pixel sampling of 7\farcs3. The rms noise level per channel is 0.45 K. To improve the S/N, we averaged the velocity channel maps over a 3$\times$3 pixel box, resulting in a final rms noise level of 0.15 K. \\
\subsection{Planck and Herschel}\label{subsec:obs_planck_herschel}
We used Planck 850 $\mu$m polarization data and Herschel 350 $\mu$m continuum data to investigate the large-scale magnetic fields and density structures surrounding L1527. The Planck 850 $\mu$m polarization data have an effective beam size of 5$^{'}$ ($\sim$0.2 pc) with a pixel spacing of 1$^{'}$ \citep{Placnk_taurus_2018XI}. The Herschel 350 $\mu$m continuum data were obtained from the Herschel Gould Belt Survey archive \citep[HGBS,][]{SPIRE_gouldbelt}. These data have an effective beam size of 25$^{\prime\prime}$ ($\sim$3500 au) with a pixel size of 10$^{\prime\prime}$. \\

\section{Results} \label{sec:results}
\subsection{Magnetic field morphology} \label{subsec:Bfield}
\begin{figure*}
    \centering
    \includegraphics[width=0.85\linewidth]{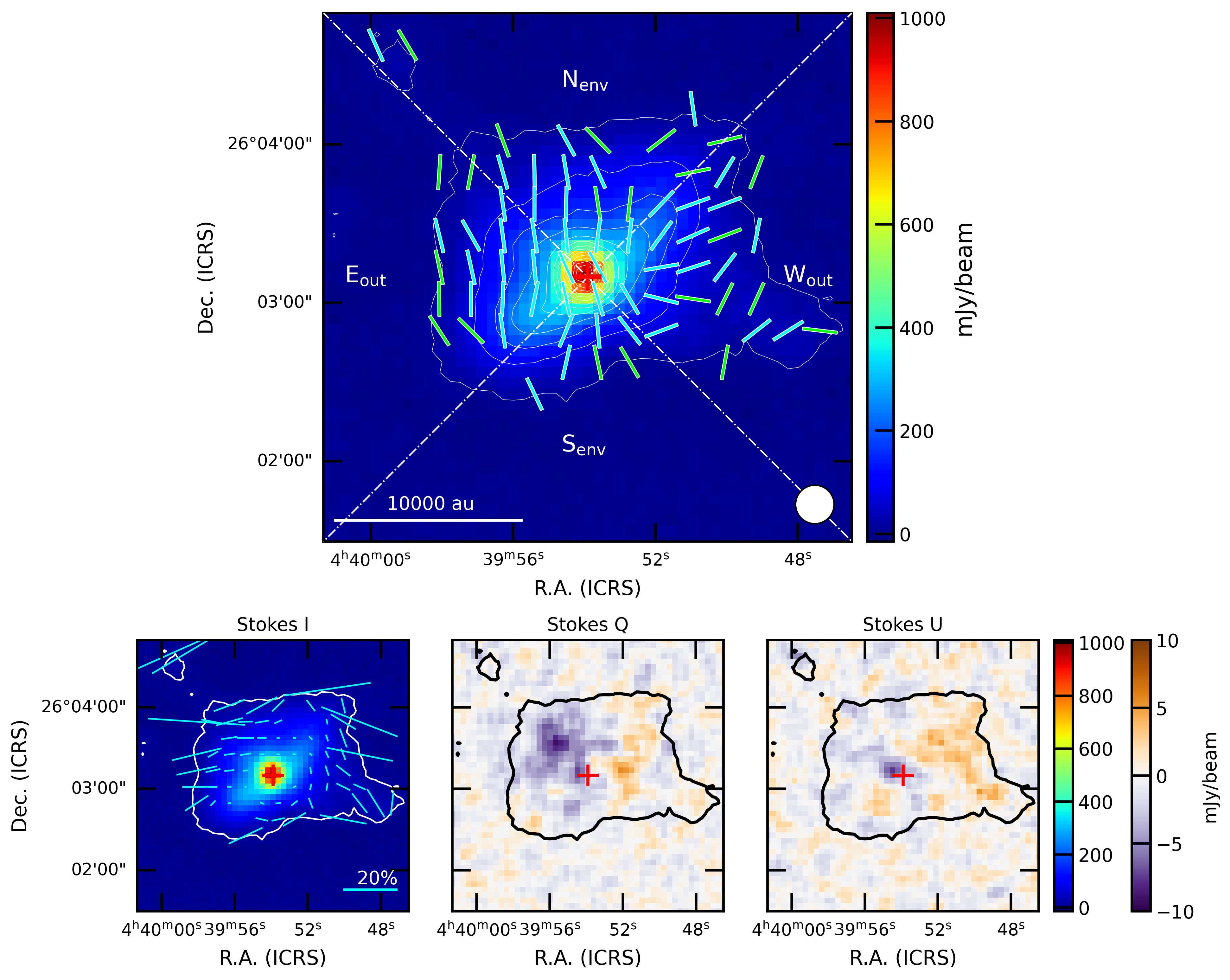}
    \caption{\textit{Top}: 850 $\mu$m intensity map overlaid with B-field orientations. The circle in the bottom-right corner indicates the effective beam size. The lowest contour corresponds to the 3$\sigma$ level with contour intervals of 20$\sigma$. The B-field orientations were obtained by rotating the polarization orientations by $90^{\circ}$ and are normalized to be the same size. Cyan segments satisfy $I/\sigma_I > 3$ and $PI/\sigma_{PI} > 3$, while yellow segments satisfy $I/\sigma_I > 3$ and $2 < PI/\sigma_{PI} \leq 3$. The regions are divided into four parts by the white dash-dotted lines based on the outflow cavity \citep[e.g.,][]{eDisk_L1527}. $\text{E}_{\text{out}}$ and $\text{W}_{\text{out}}$ correspond to the bipolar outflow regions, while $\text{N}_{\text{env}}$ and $\text{S}_{\text{env}}$ are the envelope regions. \textit{Bottom}: Stokes $I$, $Q$, and $U$ maps at 850 $\mu$m with a pixel size of $4^{\prime\prime}\times4^{\prime\prime}$ (from the left). The $I$ map is overlaid with polarization orientations scaled by polarization fraction. The $Q$ and $U$ maps are averaged over a $3\times3$ pixel box to illustrate the significance of the binned Stokes $Q$ and $U$ values used to derive the B-field orientations. The left colorbar is for the Stokes $I$, and the right is for the Stokes $Q$ and $U$.}
    \label{fig:B-vectors}
\end{figure*}

Plane of the sky magnetic field orientations are inferred by rotating the polarization angles by $90^{\circ}$, assuming that the minor axes of the dust grains are aligned with the magnetic field \citep[e.g.,][]{Lazarian_and_Hoang}. To select reliable measurements, we take the data points that satisfy the following intensity and polarization intensity thresholds: $I \, > \,3\sigma_{I}$ and $I_p \,> \, 2\sigma_{PI}$. The top panel of Figure~\ref{fig:B-vectors} shows the inferred B-field orientations overlaid on the 850 $\mu$m intensity map. Cyan segments have $I/\sigma_I > 3$ and $PI/\sigma_{PI} > 3$, while yellow segments correspond to $I/\sigma_I > 3$ and $2 < PI/\sigma_{PI} \leq 3$. Polarization orientations scaled by polarization fraction are shown in the bottom-left panel of Figure~\ref{fig:B-vectors}, while the bottom-middle and bottom-right panels present the Stokes $Q$ and $U$ maps, respectively.

\begin{figure*}
    \centering
    \includegraphics[width=0.70\linewidth]{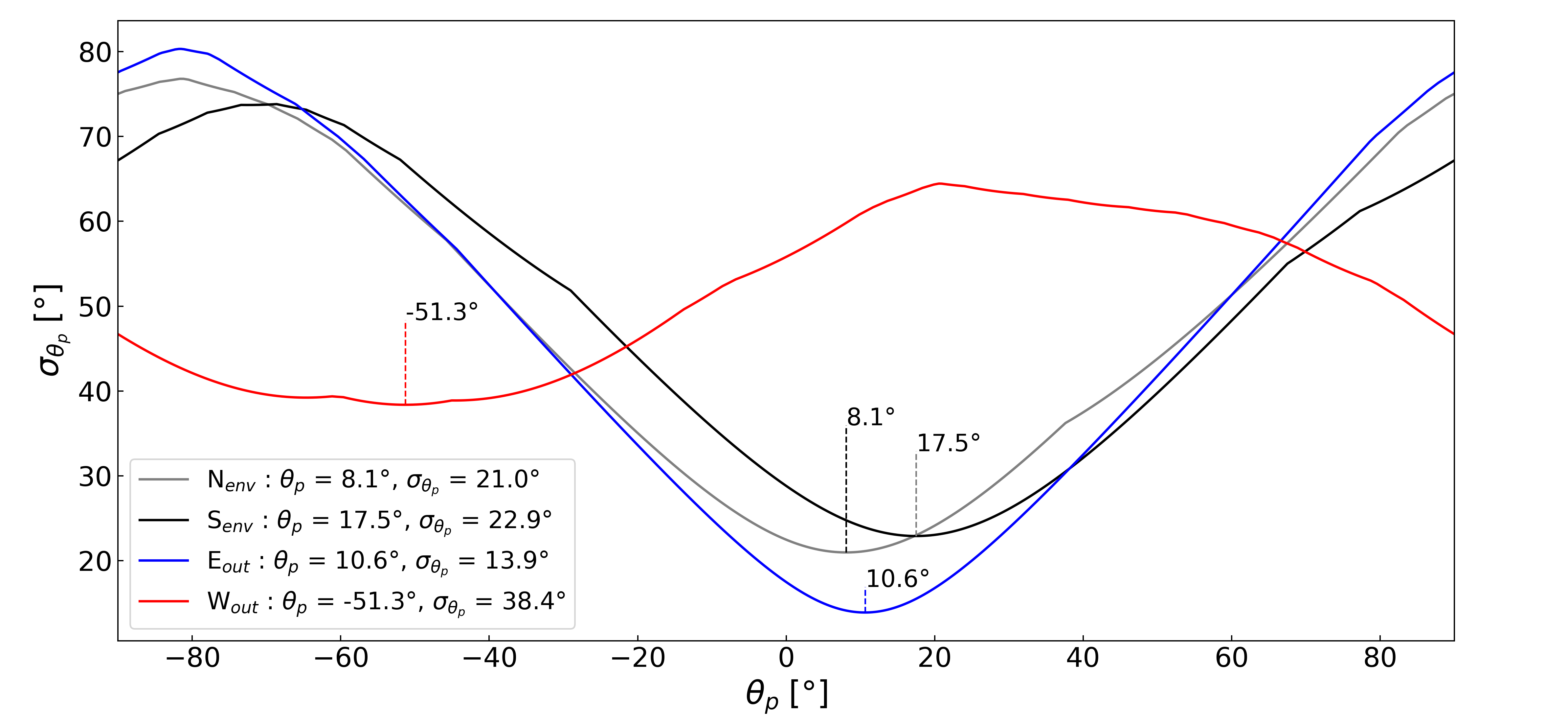}
    \caption{Results of deviation test for the four regions. Horizontal and vertical axes represent $\theta_p$ and $\sigma_{\theta_p}$, respectively.}
    \label{fig:B-vectors_polar}
\end{figure*}

\begin{figure*}
    \centering
    \includegraphics[width=0.80\linewidth]{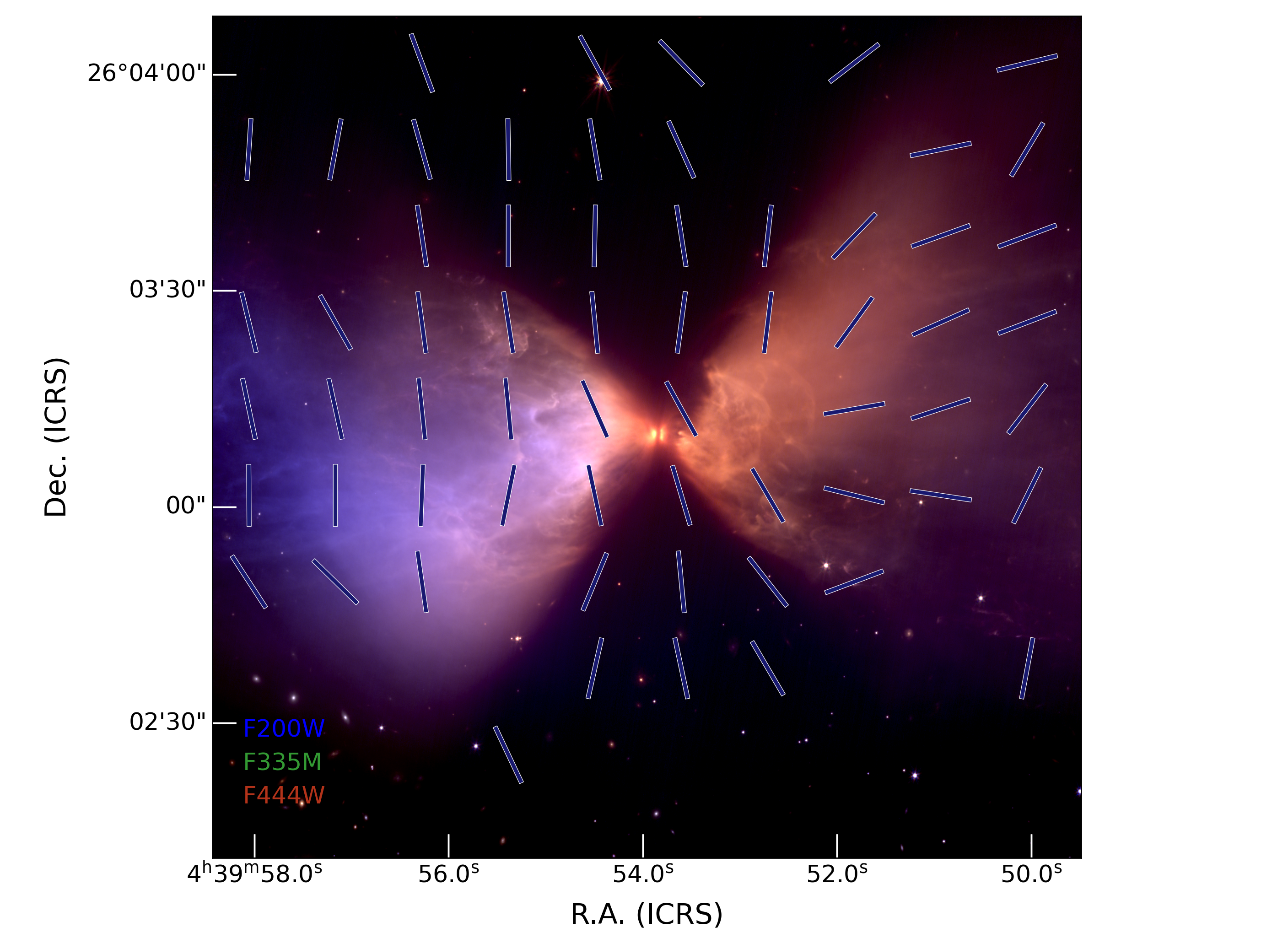}
    \caption{JWST NIRCam RGB-composited color image overlaid with the B-field orientations in Figure~\ref{fig:B-vectors}. F200W (2 $\mu$m, blue), F335M (3.3 $\mu$m, green), and F444W (4.4 $\mu$m, red) filters are combined.}
    \label{fig:B-vectors_JWST}
\end{figure*}

\begin{figure*}
    \centering
    \includegraphics[width=0.9\linewidth]{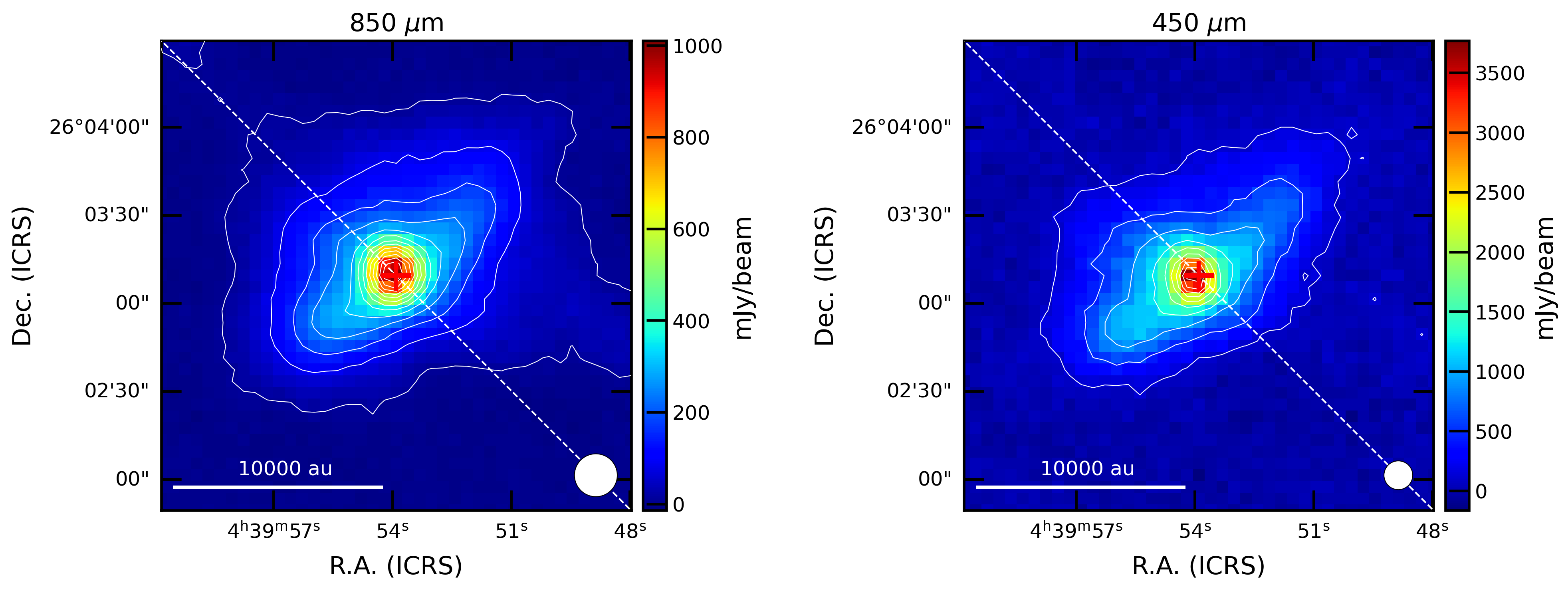}
    \caption{Intensity maps of L1527 observed with JCMT SCUBA-2/POL-2. The left and right panels show 850 $\mu$m and 450 $\mu$m intensity maps, respectively. In each panel, the red cross marks the central protostellar position, and the circle of the bottom-right corner indicates the effective beam size. The lowest contour in both images corresponds to the 3$\sigma$ level. Contour intervals are 20$\sigma$ for the 850 $\mu$m and 10$\sigma$ for the 450 $\mu$m. The white dashed line indicates the boundary between $\text{NW}_\text{core}$ and $\text{SE}_\text{core}$, where the intensity distribution appears symmetric.}
    \label{fig:intensity_fig}
\end{figure*}

As the L1527 outflow extends along the east-west direction with an opening angle of 45$^{\circ}$ \citep[e.g.,][]{eDisk_L1527}, we divided the L1527 region into four parts, as shown in Figure~\ref{fig:B-vectors} \myterm{($\text{E}_{\text{out}}$, $\text{W}_{\text{out}}$, $\text{N}_{\text{env}}$, and $\text{S}_{\text{env}}$)}. Qualitatively, the B-field orientations in the eastern outflow region \myterm{($\text{E}_{\text{out}}$)} show nearly north-south orientation, perpendicular to the outflow direction. In contrast, those in the western region \myterm{($\text{W}_{\text{out}}$)} show a pinched trend toward the center, resembling an hourglass morphology observed in various star formation regions \citep[e.g.,][]{Girart, Stephens, Maury_2018, Kwon_hourglass, Choi_HH211_hourglass}. The northern and southern envelope regions \myterm{($\text{N}_{\text{env}}$ and $\text{S}_{\text{env}}$)} also show a pinched morphology, aligned along the north-south orientation.

\myterm{To quantify the characteristics of the B-fields in each region, we introduce a deviation analysis method to describe the orientation distribution.} First, we generate a set of position angles, $\theta_{p}$, covering the range from $-90^{\circ}$ to $+90^{\circ}$ at $0.01^{\circ}$ intervals. Second, for each position angle in the set, we calculate the standard deviation:

\begin{equation}
\sigma_{\theta_p}
=
\sqrt{
\sum_{i=1}^{N}
\frac{(\theta_i-\theta_p)^2}{N}
}
\label{eq:sigma_theta}
\end{equation} 
where $\theta_{i}$ is the position angle of a B-field orientation, and $N$ is the number of B-field orientations. The value of $\theta_i-\theta_p$ is calculated as the minimum angular separation, considering the degeneracy at $\pm$ $90^{\circ}$ (e.g., the separation between $\theta_i$=$-89^{\circ}$ and $\theta_p$=$+89^{\circ}$ is $2^{\circ}$). Lastly, we adopt a minimum $\sigma_{\theta_p}$ as the angular dispersion and select the corresponding $\theta_{p}$ as a representative B-field orientation (Fig.~\ref{fig:B-vectors_polar}). \myterm{Unlike simple circular statistics, which provide only a single best-fit mean orientation and a single dispersion value from a limited number of discrete data points, this method evaluates the angular deviation over a continuous range of trial position angles ($\theta_p$).}

\myterm{$\text{W}_{\text{out}}$ stands out from the other regions in its magnetic field orientation. $\text{E}_{\text{out}}$, $\text{N}_{\text{env}}$, and $\text{S}_{\text{env}}$ have $\theta_p$ values clustered around $12^{\circ}$, with $\sigma_{\theta_p}$ values of $13.9^{\circ}$, $21.0^{\circ}$, and $22.9^{\circ}$, respectively. In contrast, $\text{W}_{\text{out}}$ has $\theta_p=-51.3^{\circ}$ and $\sigma_{\theta_p}=38.4^{\circ}$, indicating both a distinct magnetic field orientation and a much broader distribution of field angles. This difference is also reflected in the curve shapes: $\text{E}_{\text{out}}$ shows a sharp, well-defined minimum, whereas $\text{W}_{\text{out}}$ shows a relatively flat minimum extending from $-80^\circ$ to $-30^\circ$. This also illustrates an advantage of this method over simple circular statistics: the curve shape provides a visual indication of how tightly the mean orientation is constrained. The broad minimum in $\text{W}_{\text{out}}$ indicates that its mean orientation is less tightly constrained, consistent with its large angular dispersion.}

These peculiar magnetic field patterns become more evident when compared with near-infrared (NIR) observations. We retrieved JWST/NIRCam data of L1527 from the Mikulski Archive for Space Telescopes (MAST) and constructed an RGB composite image using the F200W (2~$\mu$m, blue), F335M (3.3~$\mu$m, green), and F444W (4.4~$\mu$m, red) filters to illustrate the NIR color in the outflow cavity. Figure~\ref{fig:B-vectors_JWST} shows the composite image overlaid with the magnetic field orientations. $\text{E}_{\text{out}}$ shows a bluer color and a relatively uniform magnetic field orientation that is roughly perpendicular to the outflow axis. In contrast, $\text{W}_{\text{out}}$ appears redder and exhibits a pinched magnetic field morphology that is well aligned with the outflow. These color differences imply that $\text{W}_{\text{out}}$ contains a larger amount of envelope dust that preferentially attenuates bluer (2~$\mu$m) NIR light from the cavity (Sec.~\ref{subsec:spectral_index}). These may reflect an anisotropic mass accretion history related to the magnetic fields, as discussed in Section~\ref{subsec:SF_scenario}.\\

\subsection{Intensity variations}\label{subsec:intensity}
Figure~\ref{fig:intensity_fig} shows the total intensity maps at 850 $\mu$m and 450 $\mu$m. Both maps show a similar trend, with emission extending along the northwest-southeast direction. This elongated structure may originate from an asymmetrically heated cavity wall, possibly caused by a misaligned disk relative to the outflow cavity or compact binary system \citep{Wood_2001, Loinard_2002}. \myterm{In this section, we group $\text{E}_{\text{out}}$ and $\text{S}_{\text{env}}$ into one region ($\text{SE}_\text{core}$), and $\text{W}_{\text{out}}$ and $\text{N}_{\text{env}}$ into another ($\text{NW}_\text{core}$), and compare the spatial extents of the two regions.}

\begin{figure}
    \centering
    \includegraphics[width=1\linewidth]{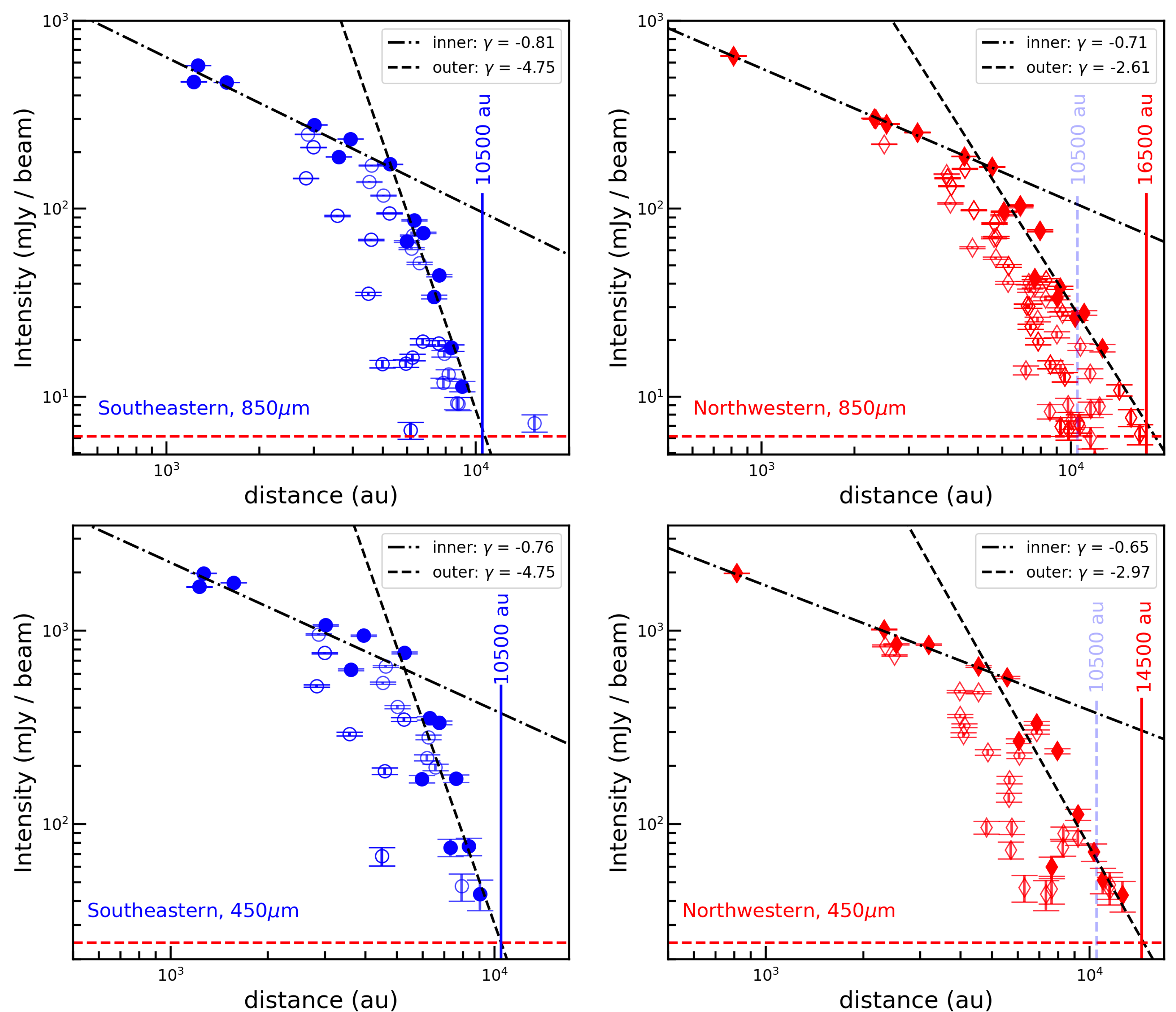}
    \caption{Broken power-law fitting results for the 850 $\mu$m (\textit{top}) and 450 $\mu$m (\textit{bottom}) intensity profiles. Blue and red dots represent the data points of the southeastern ($\text{SE}_\text{core}$) and northwestern ($\text{NW}_\text{core}$) regions, respectively. The red dashed horizontal line indicates the 5$\sigma$ noise level. The solid symbols indicate the intensity local maxima used for the broken power-law fitting. $\gamma$ denotes the index of the power-law function.}
    \label{fig:lumfit_all}
\end{figure}

To quantify how the structure extends in each region, we analyzed the intensity profiles of $\text{NW}_\text{core}$ and $\text{SE}_\text{core}$. We selected the $12^{\prime\prime}\times12^{\prime\prime}$ binned Stokes $I$ data with $I> 5\sigma_{I}$ and constructed radial intensity profiles as a function of projected distance from the center. In each region, the profile was divided into inner and outer regimes, and the local intensity maxima were identified to trace the overall intensity profile more clearly (solid symbols in Fig.~\ref{fig:lumfit_all}). These peak points were then fitted separately in the inner and outer regimes using the \texttt{curve\_fit} function from the \texttt{scipy.optimize} package, with a broken power-law function of the form: 

\begin{equation}
I(r)=
\begin{cases}
I_{\mathrm{in}} \left( \dfrac{r}{r_0} \right)^{\gamma_{\mathrm{in}}}, & \text{if } r \leq r_0 \\[8pt]
I_{\mathrm{out}} \left( \dfrac{r}{r_0} \right)^{\gamma_{\mathrm{out}}}, & \text{if } r > r_0
\end{cases},
\label{eq:broken_powerlaw}
\end{equation}
where we adopted $r_0 = 5000$ au as the radius at which the power-law slope clearly changes.

Figure~\ref{fig:lumfit_all} shows the fitting results of the 850 $\mu$m (top panels) and 450 $\mu$m (bottom panels) observations. The horizontal axis shows the projected distance from the center in au, and the vertical axis shows the intensity in mJy~beam$^{-1}$. The dash-dotted line represents the fitted function for the inner regime, while the dashed line corresponds to the outer regime. The red dashed horizontal line indicates the 5$\sigma$ noise level.

In the 850 $\mu$m intensity profiles, $\text{SE}_\text{core}$ and $\text{NW}_\text{core}$ show similar power-law indices in the inner regime. In the outer regime, however, $\text{SE}_\text{core}$ has a steeper profile, with a power-law index lower by 2.1 than that of $\text{NW}_\text{core}$. Moreover, $\text{NW}_\text{core}$ profile extends to $\sim$16500 au, whereas $\text{SE}_\text{core}$ profile is limited to $\sim$10000 au. The 450 $\mu$m intensity profiles show a similar trend. The inner regimes have comparable power-law indices, while the outer regime in $\text{SE}_\text{core}$ is steeper by about 1.8. These results suggest that, despite the overall symmetry in the intensity distribution, $\text{NW}_\text{core}$ has a more extended density structure that declines more slowly than in $\text{SE}_\text{core}$.\\

\subsection{Spectral index}\label{subsec:spectral_index}

\begin{figure*}
    \centering
    \includegraphics[width=1\linewidth]{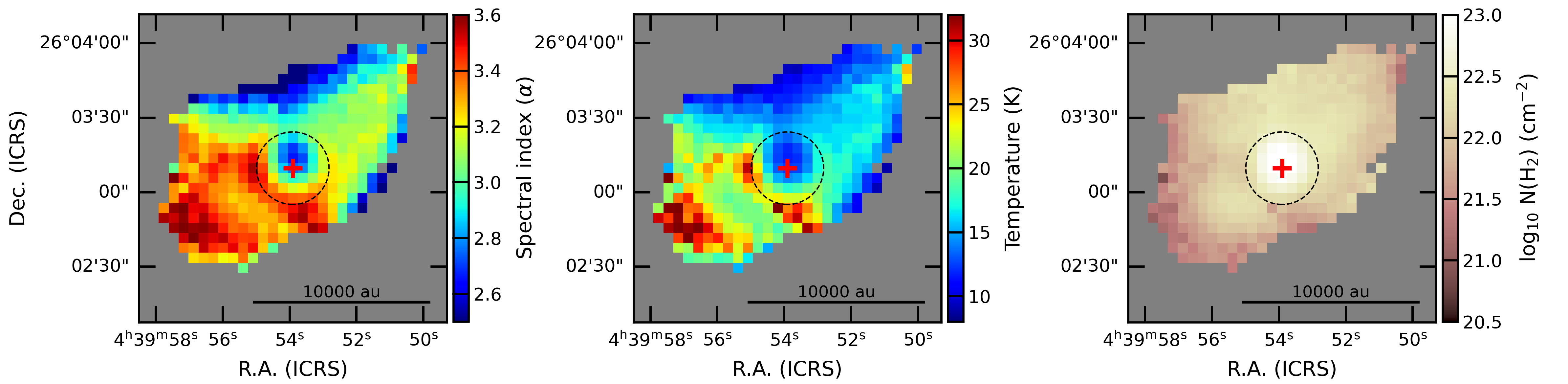}
    \caption{\textit{Left}: Spectral index map derived from the 450 $\mu$m and 850 $\mu$m data. \textit{Middle}: Dust temperature map calculated from the spectral index under the assumption of optically thin emission. \textit{Right}: Molecular hydrogen column density map derived from the intensity and temperature maps. The central circle in all panels indicates a diameter of 29.2$^{\prime\prime}$ (twice the beam diameter), marking the region potentially affected by optically thick emission from the central protostellar disk.}
    \label{fig:spec_index}
\end{figure*}

The spectral index provides the physical properties of dust thermal emission, such as dust temperature and density \citep[e.g.,][]{Kirk_taurus}. It is defined as 

\begin{equation}
\alpha = \frac{\log\left(F_{\nu_2}/F_{\nu_1}\right)}{\log\left(\nu_2/\nu_1\right)},
\label{eq:spectral_index}
\end{equation}
where $F_{\nu_1}$ and $F_{\nu_2}$ are the monochromatic fluxes at frequencies $\nu_1$ and $\nu_2$, respectively. In the (sub)millimeter, the dust opacity is generally assumed to follow a power-law dependence on frequency: $\kappa_{\nu} \propto \nu^{\beta}$. The power-law index $\beta$, known as the dust opacity spectral index, depends on dust properties such as grain size distribution and composition \citep[e.g.,][]{Hildebrand_dust, Draine_beta}. Assuming that the submillimeter dust thermal emission is optically thin---which is typically the case in the envelopes of low-mass star-forming cores---the intensity ratio between the two frequencies, $\nu_1$ and $\nu_2$, can be expressed as

\begin{equation}
\frac{I_{\nu_2}}{I_{\nu_1}}
=
\left(\frac{\nu_2}{\nu_1}\right)^{\beta+3}
\frac{\exp(h\nu_1/kT_d)-1}{\exp(h\nu_2/kT_d)-1},
\label{eq:intensity_ratio}
\end{equation}
where $h$ is the Planck constant, $k$ is the Boltzmann constant, and $T_d$ is the the line-of-sight averaged dust temperature. Therefore, if the beam sizes at the two frequencies are identical, the spectral index can be calculated directly from the intensity ratio:

\begin{equation}
\alpha
=
\frac{\log\left(F_{\nu_2}/F_{\nu_1}\right)}{\log\left(\nu_2/\nu_1\right)}
=
\frac{\log\left(I_{\nu_2}/I_{\nu_1}\right)}{\log\left(\nu_2/\nu_1\right)}.
\label{eq:alpha_definition}
\end{equation} 
The JCMT POL-2/SCUBA-2 instrument provides intensity data at both 450 $\mu$m and 850 $\mu$m simultaneously. Prior to the spectral index calculation, the beam sizes of the intensity maps were matched. The effective Gaussian beam sizes are 9\farcs8 at 450 $\mu$m and 14\farcs6 at 850 $\mu$m. To achieve a common resolution, the 450 $\mu$m intensity map was convolved with a Gaussian kernel corresponding to the difference in beam sizes. We then selected pixels with $I\,>\,4\sigma_{I}$ in both wavelengths and calculated the spectral index.

The spectral index varies depending on the underlying physical conditions. If the dust opacity spectral index ($\beta$) is constant, higher dust temperatures lead to higher spectral indices. If the temperature is constant, larger $\beta$ values result in higher spectral indices; in particular, grain growth can significantly reduce $\beta$ \citep[e.g.,][]{Draine_beta, Kwon_dustgrainsize}. In compact, high-density regions where the optically thin assumption breaks down (e.g., in protostellar disks), the spectral index can be significantly reduced. For instance, when the emission becomes optically thick, the spectral index can approach $\alpha \sim 2$ and the scattering opacity may be significant \citep[e.g.,][]{Birnstiel_scatt}. In addition, if observations at two frequencies are taken at significantly different epochs, the derived spectral index distribution may be affected by protostellar variability.

In this study, we combined data sets obtained in 2017 and 2023. To verify that the combination is valid and to assess the effect of variability, we reduced the two data sets separately. We found that the peak intensities at the two epochs differ by only $\sim5$\% at both 850 and 450 $\mu$m, and we also confirmed that the resulting spectral index distributions are consistent with each other.

We assumed optically thin emission and adopted a uniform dust opacity spectral index of $\beta \approx 2$ for the embedded core \citep{Draine_beta_2}. The dust temperature is then derived numerically from the spectral index pixel by pixel using Equation~\ref{eq:intensity_ratio}, and the molecular hydrogen column density is calculated as

\begin{equation}
N(\mathrm{H}_2)=\frac{I_{\nu}}{\mu m_{H}\kappa_{\nu}B_{\nu}(T_d)},
\label{eq:column_density}
\end{equation}
where $m_{H}$ is the hydrogen mass and $\mu = 2.86$ is the mean molecular weight per hydrogen molecule, assuming that hydrogen accounts for approximately 70\% of the total gas mass. We adopted $\kappa_{\nu} = 0.012\,\mathrm{cm^2\,g^{-1}}$ at 850 $\mu$m, assuming a gas-to-dust mass ratio of 100 \citep{kappa_Ref}.

L1527 shows a central hole in the spectral index map, which can be understood by optically thick emission from the protostellar disk (Fig.~\ref{fig:spec_index}). In the other regions, the spectral index exhibits distinct spatial variations: $\text{E}_{\text{out}}$ and $\text{S}_{\text{env}}$ have higher values, $\text{W}_{\text{out}}$ shows intermediate values, and $\text{N}_{\text{env}}$ has the lowest values. The corresponding temperature map shows the similar pattern. Considering the symmetry in the intensity maps (Fig.~\ref{fig:intensity_fig}), this temperature distribution suggests higher densities in $\text{W}_\text{out}$ than in $\text{E}_\text{out}$, and in $\text{N}_\text{env}$ than in $\text{S}_\text{env}$. This trend is also clearly seen in the $N(\text{H}_2)$ column density map (right panel of Fig.~\ref{fig:spec_index}).

\begin{figure*}
    \centering
    \includegraphics[width=0.9\linewidth]{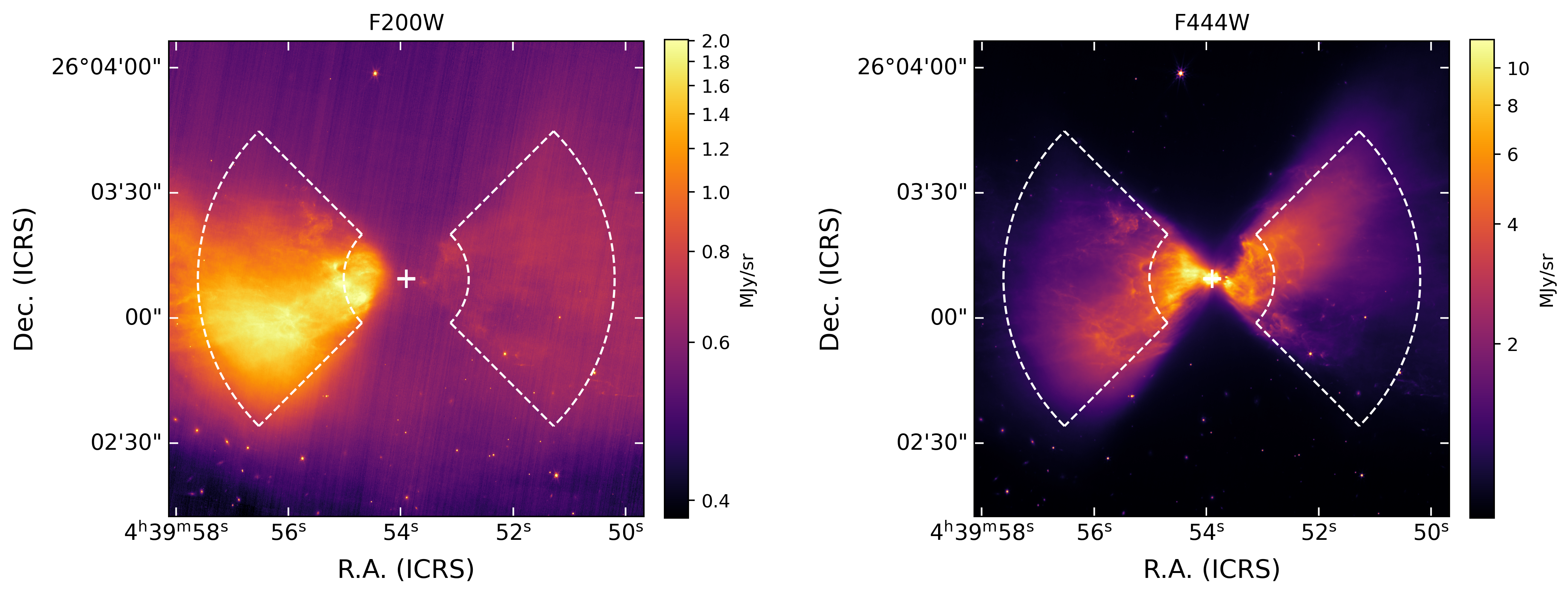}
    \caption{\textit{Left}: JWST NIRCam 2~$\mu$m intensity map. \textit{Right}: JWST NIRCam 4.4~$\mu$m intensity map. Both images are shown in the same intensity unit, MJy~sr$^{-1}$. The white dashed regions indicate $\text{E}_{\text{out}}$ and $\text{W}_{\text{out}}$, excluding the optically thick central hole in Figure~\ref{fig:spec_index}. The cross marks the location of the L1527 protostar. The mean intensities within $\text{E}_{\text{out}}$ and $\text{W}_{\text{out}}$ are 1.16~MJy~sr$^{-1}$ and 0.66~MJy~sr$^{-1}$ at 2~$\mu$m, and 1.93~MJy~sr$^{-1}$ and 1.80~MJy~sr$^{-1}$ at 4.4~$\mu$m, respectively.}
    \label{fig:jwst_quantify}
\end{figure*}

\myterm{We further tested whether the asymmetric distribution of the spectral index can be explained by variations in $\beta$ within the L1527 core, rather than by changes in the asymmetric temperature distribution (see App.~\ref{app:beta_var}). We treated the temperature in Equation~\ref{eq:intensity_ratio} as a fixed parameter and examined the required variation in $\beta$ over a temperature range from 10~K to 35~K. We found that, across the entire temperature range, variations in $\beta$ of at least 0.6 is required to reproduce the observed results, which is not generally expected in core structures of a few tens thousands au scales. In addition, for all temperatures considered, the derived values of $\beta$ in some regions would have to be either significantly low ($\beta<1.4$) or high ($\beta>2.4$). Therefore, we conclude that the asymmetric spectral index distribution observed in L1527 primarily reflects intrinsic variations in temperature and column density, rather than variations in dust properties.}

We measured the mean temperature and molecular hydrogen column density of four regions excluding the optically thick center. \myterm{The mean temperature uncertainty, $\delta T_d$, is calculated as}

\begin{equation}
\begin{aligned}
&\frac{1}{T_d^2}
\left(
\frac{\frac{h\nu_1}{k} e^{h\nu_1/kT_d}}
{e^{h\nu_1/kT_d}-1}
-
\frac{\frac{h\nu_2}{k} e^{h\nu_2/kT_d}}
{e^{h\nu_2/kT_d}-1}
\right)^2
\left(\frac{\delta T_d}{T_d}\right)^2 \\
&=
\left(\frac{\delta I_{\nu_1}}{I_{\nu_1}}\right)^2
+
\left(\frac{\delta I_{\nu_2}}{I_{\nu_2}}\right)^2
+
\left(
\delta\beta
\ln\left(\frac{\nu_1}{\nu_2}\right)
\right)^2,
\end{aligned}
\label{eq:temp_error}
\end{equation}
\myterm{where $I$ is the mean intensity, $\delta I$ is the rms noise level, and $\delta \beta$ is the systemic variation of the dust opacity spectral index, which we account for $\pm0.2$. Based on the temperature uncertainty, we also derived the molecular hydrogen column density uncertainty as:}

\begin{equation}
\begin{gathered}
\left( \frac{\delta N(\mathrm{H}_2)}{N(\mathrm{H}_2)} \right)^2
=
\left(\frac{\delta I_{\nu}}{I_{\nu}}\right)^2
+
\left(\frac{\delta \kappa_{\nu}} {\kappa_{\nu}}\right)^2 \\
+
\left(
\frac{\partial \ln B_\nu}{\partial \ln T_d}
\frac{\delta T_d}{T_d}
\right)^2,
\end{gathered}
\label{eq:column_error}
\end{equation}
\myterm{where we adopted the dust opacity uncertainty as 50\%. $\text{E}_{\text{out}}$ and $\text{S}_{\text{env}}$ show higher mean temperatures of $22.5\pm7.0$~K and $23.0\pm7.6$~K, respectively, whereas $\text{W}_{\text{out}}$ and $\text{N}_{\text{env}}$ have $15.9\pm3.1$~K and $13.1\pm2.0$~K, respectively. The molecular hydrogen column densities in $\text{E}_{\text{out}}$ and $\text{S}_{\text{env}}$ are $8.2\pm5.1\times 10^{21}$~cm$^{-2}$ and $6.9\pm4.5\times 10^{21}$~cm$^{-2}$, respectively, while $\text{W}_{\text{out}}$ and $\text{N}_{\text{env}}$ have mean column densities of  $12.5\pm7\times 10^{21}$~cm$^{-2}$ and $17.2\pm10\times 10^{21}$~cm$^{-2}$, respectively. Based on the column density estimations, we derived the envelope mass by assuming that $N(\mathrm{H}_2)$ in the optically thick region is equal to the boundary value of the central region (black dashed circle), $\sim25\times10^{21}$~cm$^{-2}$, although the true column density is likely larger. This yields a lower limit to the envelope mass of $0.75 \pm 0.45\,M_{\odot}$.}

\myterm{Using the column density difference between $\text{E}_{\text{out}}$ and $\text{W}_{\text{out}}$ regions inferred from the sub-mm spectral index, we analyzed whether this difference in the envelope mass distribution can account for the color difference of the outflow cavity seen in the NIR image in Figure~\ref{fig:B-vectors_JWST}. We assumed that the intrinsic NIR intensity inside the outflow cavity is the same on $\text{E}_{\text{out}}$ and $\text{W}_{\text{out}}$, for the background of the envelope material. Under this assumption, the ratio of the NIR intensities between $\text{E}_{\text{out}}$ and $\text{W}_{\text{out}}$ at a given frequency can be expressed in terms of the dust opacity and the column densities derived from the submillimeter observations:}
\begin{equation}
\begin{gathered}
R_{\nu} = \frac{I_0 e^{-\tau_{\nu,E}/2}}{I_0 e^{-\tau_{\nu,W}/2}}=\exp\left({\kappa_{\nu}} \mu m_{\mathrm{H}}\frac{\Delta N(\mathrm{H}_2)}{2} \right)
\end{gathered}
\label{eq:color_diff_obs}
\end{equation}
\myterm{where $\Delta N(\mathrm{H}_2) = N_W(\mathrm{H}_2)-N_E(\mathrm{H}_2)=4.3\times 10^{21}\,\text{cm}^{-2}$. Here, we assumed that half of the observed column density lies in the foreground and contributes to the extinction of NIR light from the outflow cavity, based on the nearly edge-on geometry of the system. For the dust opacity in the NIR regime, we adopted $\kappa_{2\mu{\rm m}} = 57.48~\mathrm{cm^2~g^{-1}}$ and $\kappa_{4.4\mu{\rm m}} = 17.22~\mathrm{cm^2~g^{-1}}$ from \citet{Ossenkopf_Henning}. This gives expected intensity ratios of $R_{2\mu{\rm m}}=1.85$ and $R_{4.4\mu{\rm m}}=1.20$. To compare these values with the JWST/NIRCam observations, we measured the mean intensities within the white dashed regions corresponding to $\text{E}_{\text{out}}$ and $\text{W}_{\text{out}}$ (Fig.~\ref{fig:jwst_quantify}). The resulting eastern-to-western intensity ratios are 1.76 at 2~$\mu$m and 1.07 at 4.4~$\mu$m, in broad agreement with the expected values from the submillimeter observations. The corresponding mean intensities measured within $\text{E}_{\text{out}}$ and $\text{W}_{\text{out}}$ are listed in the caption of Figure~\ref{fig:jwst_quantify}. These results suggest that the NIR color difference can be plausibly explained by different levels of foreground extinction by envelope dust, although substantial uncertainties remain in the column density estimates and in the fraction of material located in the foreground.} \\


\section{Analysis and Discussion} \label{sec:discussion}
\subsection{Magnetic field strength} \label{subsec:DCF}

To better understand the role of magnetic fields in the L1527 system, we measured the magnetic field strength in each region defined in Section~\ref{subsec:Bfield}. The magnetic field strength projected onto the plane of the sky can be estimated using the Davis--Chandrasekhar--Fermi (DCF) method \citep{Davis_DCF,CF_DCF}. This method estimates the plane-of-sky magnetic field strength from the ratio of the turbulent velocity dispersion to the dispersion in B-field orientations at a given gas density. The magnetic field strength is then given by: 

\begin{equation}
B_{\mathrm{POS}}
=
\mathcal{F}\sqrt{4\pi\rho}\frac{\sigma_{v,\mathrm{turb}}}{\delta\phi}
\approx
9.3\,\sqrt{n(\mathrm{H}_2)}\,\frac{\Delta V}{\delta\phi}
\label{eq:DCF}
\end{equation}
where $\mathcal{F}$ is a factor of the order of unity, and we adopt $\mathcal{F}\text{ = }0.5$ in this study \citep{Ostriker_2011}. $\rho$ is the gas density, $\sigma_{v,\mathrm{turb}}$ is the turbulent velocity dispersion, and $\delta \phi$ is the dispersion in B-field orientations in degrees. $n(\mathrm{H}_2)$ and $\Delta V$ are the molecular hydrogen number density and the FWHM line width of the turbulent component, in units of $\mathrm{cm}^{-3}$ and $\mathrm{km\,s^{-1}}$, respectively. The resulting magnetic field strength is given in $\mu\mathrm{G}$ \citep{crutcher_dcf}.

We applied the DCF method separately to each of the four regions defined in Section~\ref{subsec:Bfield}. We first estimated $\delta\phi$ in each region using the unsharp masking technique \citep[e.g.,][]{Pattle_BISTRO}. We used the magnetic field orientations shown in Figure~\ref{fig:B-vectors}, whose sampling size is $12^{\prime\prime} \times 12^{\prime\prime}$, comparable to the beam size. \myterm{We estimated the background field (i.e., mean fields without turbulent components) orientation, $\langle \theta \rangle$, by smoothing the polarization orientations. To avoid the ambiguity associated with directly averaging polarization angles, we first converted the polarization angle into normalized Stokes parameters, defined as}
\begin{equation}
Q_{\rm norm} \equiv \cos(2\chi), \quad
U_{\rm norm} \equiv \sin(2\chi),
\label{eq:norm_QU}
\end{equation}
\myterm{where $\chi$ is the polarization angle. We then smoothed the $Q_{\rm norm}$ and $U_{\rm norm}$ maps to an effective beam size of 36\farcs5, which is 2.5 times larger than the observed beam size of 14\farcs6. The background orientation was then derived from the smoothed maps.}

\begin{figure*}
    \centering
    \includegraphics[width=1\linewidth]{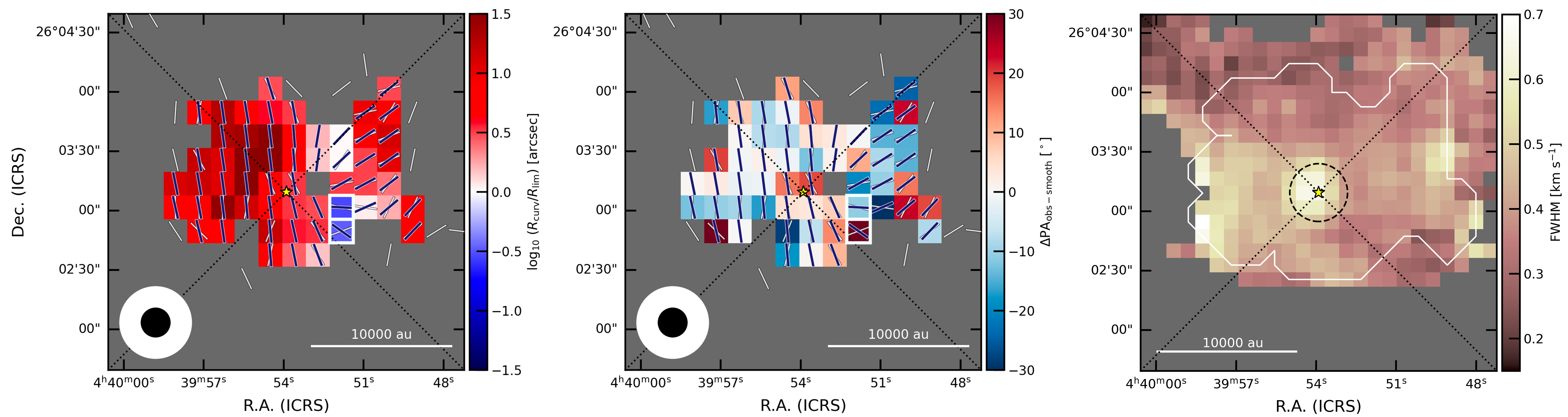}
    \caption{\textit{Left}: $\log_{10} (R_{curv}/R_{lim})$ map derived from the smoothed B-fields orientations. White boxes indicates the pixels whose $R_{curv}$ is smaller than $R_{lim}$. Blue and white B-field segments indicate the background and observed B-field orientations. White and black circle in the bottom-left corner are the observed and smoothed beam sizes, respectively. Dotted lines indicate the boundaries between regions defined in Section~\ref{subsec:Bfield}. \textit{Middle}: Angular difference map between the observed and background magnetic field orientations. \textit{Right}: FWHM map of the $\mathrm{C}^{18}\mathrm{O}$ ($J=3\text{--}2$) emission line derived from 1D Gaussian fitting. The white solid line shows the regional mask where the angular dispersion was estimated. The black dashed circle indicates the central region where the line emission may be optically thick and affected by rotational motion.}
    \label{fig:angular_dispersion_FWHM}
\end{figure*}

\myterm{The unsharp masking technique estimates the background field by smoothing the observed field orientations. For this approach to be valid, the smoothing scale should be smaller than the local radius of curvature, which represents the characteristic scale of systematic variations in the background field. To check this criterion, we estimated the local radius of curvature, $R_{\rm curv}$, following \citet{Koch_2012}:}
\begin{equation}
\frac{1}{R_{\rm curv}}
=
\frac{2}{d}
\sin\left(\frac{\Delta {\rm PA}}{2}\right),
\label{eq:radius_curvature}
\end{equation}
\myterm{where $d$ is the separation between the central pixel and each neighboring pixel, and $\Delta{\rm PA}$ is the corresponding difference in the background field position angle.}

\myterm{For each pixel, we considered neighboring pixels within a $3\times3$ box and computed the mean radius of curvature. This calculation was performed only when at least four neighboring pixels were valid. We then adopted $R_{\rm lim}=73^{\prime\prime}$, corresponding to twice the adopted smoothing scale, as a conservative threshold. Pixels with $R_{\rm curv}<R_{\rm lim}$ were regarded as regions where the unsharp masking technique may not reliably separate the ordered field from local angular variations, and were excluded from the subsequent DCF analysis.}

\myterm{The left panel of Figure~\ref{fig:angular_dispersion_FWHM} shows this validity test. The white segments indicate the observed magnetic field orientations, while the blue segments represent the smoothed background magnetic field orientations. The color scale shows $\log_{10}(R_{\rm curv}/R_{\rm lim})$. Overall, the radius of curvature decreases in the intervening region between the eastern and western region. In particular, two pixels in $\text{W}_{\text{out}}$, marked by the white boxes, show $R_{\rm curv}<R_{\rm lim}$ and were therefore excluded from the subsequent calculations. The middle panel of Figure~\ref{fig:angular_dispersion_FWHM} shows the angular differences between the background field and the observed magnetic fields.}

\myterm{Using only pixels that satisfy $R_{\rm curv}>R_{\rm lim}$, we evaluated the magnetic field dispersion in each of the four regions from the difference between the observed magnetic field orientation and the background-field orientation:}
\begin{equation}
\delta\phi
=
\sqrt{
\frac{1}{N}
\sum_{i=1}^{N}
\left(\theta_i - \langle \theta \rangle_i\right)^2
},
\label{eq:delta_phi}
\end{equation}
\myterm{where $\theta_i$ is the observed magnetic field position angle in the $i$th pixel, $\langle \theta \rangle_i$ is the corresponding background field orientation derived from the smoothed map, and $N$ is the number of valid pixels within each region (Tab.~\ref{tab:region_properties}).}
\begin{table*}
\centering
\caption{Regional Magnetic Field Strength Measures}
\label{tab:region_properties}
\scriptsize
\begin{tabular*}{\textwidth}{@{\extracolsep{\fill}} lcccccccc}
\hline
Region &
$N(\mathrm{H}_2)$ &
$n(\mathrm{H}_2)$ &
$\delta \phi$ &
$\Delta V$ &
$B_{\rm POS}$ &
$\lambda$ &
$v_{\rm A}$ &
$\mathcal{M}_{A}$ \\
&
($10^{21}$ cm$^{-2}$) &
($10^{4}$ cm$^{-3}$) &
(deg) &
(km s$^{-1}$) &
($\mu$G) &
&
(km s$^{-1}$) &
\\
\hline
\hline
$\text{E}_{\text{out}}$  & 8.2 & 3.2 & 12.3 & 0.45 & $61 \pm 26$ & $1.01 \pm 0.75$ & $0.44 \pm 0.23$ & $0.43 \pm 0.08$ \\
$\text{W}_{\text{out}}$  & 12.5 & 4.9 & 17.2 & 0.36 & $43 \pm 17$ & $2.2 \pm 0.71$ & $0.25 \pm 0.13$ & $0.60 \pm 0.12$  \\
$\text{N}_{\text{env}}$ & 17.2 & 6.8 & 11.9 & 0.32 & $65 \pm 26$ & $2.0 \pm 0.70$ & $0.33 \pm 0.16$ & $0.42 \pm 0.08$  \\
$\text{S}_{\text{env}}$ & 6.9 & 2.7 & 14.5 & 0.35 & $37 \pm 16$ & $1.42 \pm 0.80$ & $0.29 \pm 0.16$ & $0.51 \pm 0.10$  \\
\hline
\end{tabular*}
\end{table*}

To measure the velocity dispersion, we used $\mathrm{C}^{18}\mathrm{O}$ ($J=3$--$2$) line data obtained with the JCMT HARP/ACSIS instrument (Section~\ref{subsec:obs_HARP}). For each pixel, we performed 1D Gaussian fitting using the \texttt{Gaussian1D} model in the \texttt{astropy.modeling} package and measured the FWHM of the $\mathrm{C}^{18}\mathrm{O}$ line (right panel of Fig.~\ref{fig:angular_dispersion_FWHM}). During the fitting procedure, pixels with a Gaussian peak below the $3\sigma$ noise level ($\sim$0.45 K) and a Gaussian FWHM smaller than 0.15 km~s$^{-1}$ were excluded to remove spurious features caused by noise. We also excluded the central region, where the line emission may be optically thick and the gas motion may be affected by rotation (black dashed circle). Finally, to match the region over which the angular dispersion was measured, we applied a mask to the FWHM map (white solid line in the right panel of Fig.~\ref{fig:angular_dispersion_FWHM}) and calculated the mean velocity dispersion in each region. The non-thermal turbulent velocity dispersion was then derived using the following equation:
\begin{equation}
\sigma_{v,\mathrm{turb}}^2
=
\sigma_{v,\mathrm{C^{18}O}}^2
-
\frac{kT_{gas}}{m_{\mathrm{C^{18}O}}},
\label{eq:sigma_turb}
\end{equation}
where $\sigma_{v,\mathrm{C^{18}O}}$ is the total velocity dispersion measured from the $\mathrm{C}^{18}\mathrm{O}$ line fitting, $T_{gas}$ is the gas temperature, and $m_{\mathrm{C^{18}O}}$ is the molecular mass of $\mathrm{C}^{18}\mathrm{O}$, corresponding to 30 amu. The gas temperature was adopted from the mean dust temperature in each region (Sec.~\ref{subsec:spectral_index}), assuming thermal equilibrium between gas and dust. 

\myterm{In this analysis, we assume that the $\mathrm{C}^{18}\mathrm{O}$ emission originates primarily from the same envelope material traced by the polarized 850~$\mu$m dust thermal emission. This assumption is justified as a first-order approximation, since the $\mathrm{C}^{18}\mathrm{O}$ moment 0 map shows a morphology broadly consistent with the JCMT and Herschel dust continuum maps, which mainly trace the cold dusty envelope (see Figs.~\ref{fig:intensity_fig}, \ref{fig:Herschel_Plank_JCMT}, and \ref{fig:PV_HARP}). Nevertheless, the $\mathrm{C}^{18}\mathrm{O}$ emission in these regions may include contributions from gas in the outflow cavity walls, which could systematically broaden the measured line widths and thus lead to an overestimation of $\Delta V$.}
\myterm{Conversely, outflow dynamics may also contribute to the measured angular dispersion, particularly in $\text{W}_{\text{out}}$, where the magnetic field vectors are largely aligned with the outflow axis. If the outflow perturbs the magnetic fields in the surrounding envelope, part of the observed $\delta\phi$ in $\text{W}_{\text{out}}$ may reflect outflow-driven field disturbances rather than turbulent perturbations alone. This would increase the measured angular dispersion and could lead to an underestimation of $B_{\rm POS}$. Therefore, the effects of outflow contamination on the DCF field strengths are not straightforward: broadened line widths in the outflow regions may increase the inferred field strength, whereas enhanced angular dispersions, especially in $\text{W}_{\text{out}}$, may decrease it.}

The molecular hydrogen number density was calculated from the column density derived in Section~\ref{subsec:spectral_index} and the line-of-sight depth $W$, 
\begin{equation}
n(\mathrm{H}_2)
=
\frac{N(\mathrm{H}_2)}{W}.
\label{eq:number_density}
\end{equation}
We adopted $W = 17000$ au, corresponding to the projected physical scale of the L1527 core ($\sim2^{\prime}$).

Finally, the plane-of-sky magnetic field strength ($B_\mathrm{POS}$) was derived by combining the magnetic field angular dispersion, the FWHM of the turbulent velocity, and the molecular hydrogen number density (Tab.~\ref{tab:region_properties}). 

For error calculation, we adopted the derived uncertainty of column density from Equation~\ref{eq:column_error}, and assumed 20\% uncertainty in both $\Delta V$ and $\delta\phi$ estimations. These results in a relative uncertainty in the magnetic field strength of $\sim$40\%. 

Based on the plane-of-sky magnetic field strength, we also estimated the mass-to-flux ratio ($\lambda$) relative to the critical value, which indicates whether a region is magnetically supported or dominated by gravity \citep{crutcher_dcf}:

\begin{equation}
\lambda
=
\frac{(M/\Phi)_{\mathrm{obs}}}{(M/\Phi)_{\mathrm{crit}}}
=
7.6 \times 10^{-21}\frac{N(\mathrm{H}_2) \, (\mathrm{cm}^{-2})}{B_\mathrm{POS} \, (\mathrm{\mu G})}.
\label{eq:mass_to_flux}
\end{equation}
The uncertainty in $\lambda$ was derived from the previously estimated uncertainty in the magnetic field strength and the column density. \myterm{The mass-to-flux ratio is commonly used to assess whether a whole core structure is magnetically subcritical or supercritical. Here, however, we apply this quantity to subregions within the core in order to compare relative variations in magnetic support. Therefore, the regional $\lambda$ values should be interpreted as relative proxies rather than as definitive indicators of whether each subregion is magnetically subcritical or supercritical. Interestingly, $\text{W}_{\text{out}}$ and $\text{N}_{\text{env}}$ exhibit $\lambda\approx2$, indicating that these regions are relatively less magnetically supported (Tab.~\ref{tab:region_properties}). This is broadly consistent with the expected conditions in a collapsing protostellar envelope. In contrast, $\text{E}_{\text{out}}$ and $\text{S}_{\text{env}}$ have $\lambda = 1$ and $\lambda = 1.4$, respectively, suggesting that the magnetic fields may play a relatively larger role in regulating mass accretion against gravity.}

In addition, we estimated the Alfv\'en velocity and Alfv\'en Mach number,
\begin{equation}
v_A
=
\frac{B_{\mathrm{POS}}}{\sqrt{4\pi \mu m_H n(\mathrm{H}_2)}},
\label{eq:alfven_velocity}
\end{equation}

\begin{equation}
\mathcal{M}_{A}
=
\frac{\sigma_{v,\mathrm{turb}}}{v_A}.
\label{eq:alfven_mach}
\end{equation}
Using Equations~\ref{eq:DCF}, \ref{eq:alfven_velocity}, and \ref{eq:alfven_mach}, the $v_A$ and $\mathcal{M}_{A}$ are found to depend on $\Delta V/\delta\phi$ and $\delta\phi$, respectively \citep[e.g.,][]{Cortes_2025}. The resulting uncertainties are 28\% for $v_A$ and 20\% for $\mathcal{M}_{A}$. Note that these equations include only the plane-of-sky component of the magnetic field, and the derived Alfv\'en Mach number should be regarded as an upper limit, since the true magnetic field strength may be larger when the line-of-sight component is taken into account. \myterm{All regions show clearly sub-Alfv\'enic conditions ($\mathcal{M}_{A}<0.6$). These results suggest that magnetic fields dominate over turbulence throughout the core. In particular, the magnetic fields in $\text{E}_{\text{out}}$ and $\text{S}_{\text{env}}$ are likely to play a dynamically important role in the star formation process, as indicated by their low mass-to-flux ratios and sub-Alfv\'enic Mach numbers.}\\

\subsection{Large scale structures and magnetic fields} \label{subsec:Planck_Herschel}
To understand the distinctive features, we further examined the surrounding large-scale structures and magnetic field configurations. We used the Herschel/SPIRE 350 $\mu$m dust continuum data from the Herschel Gould Belt Survey \citep[HGBS;][]{HGBS} to investigate the large-scale structure (beam size $\sim3500$ au), and the Planck 850 $\mu$m polarization data to trace the large-scale magnetic field morphology (beam size $\sim0.2$ pc).

\begin{figure}
    \centering
    \includegraphics[width=1\linewidth]{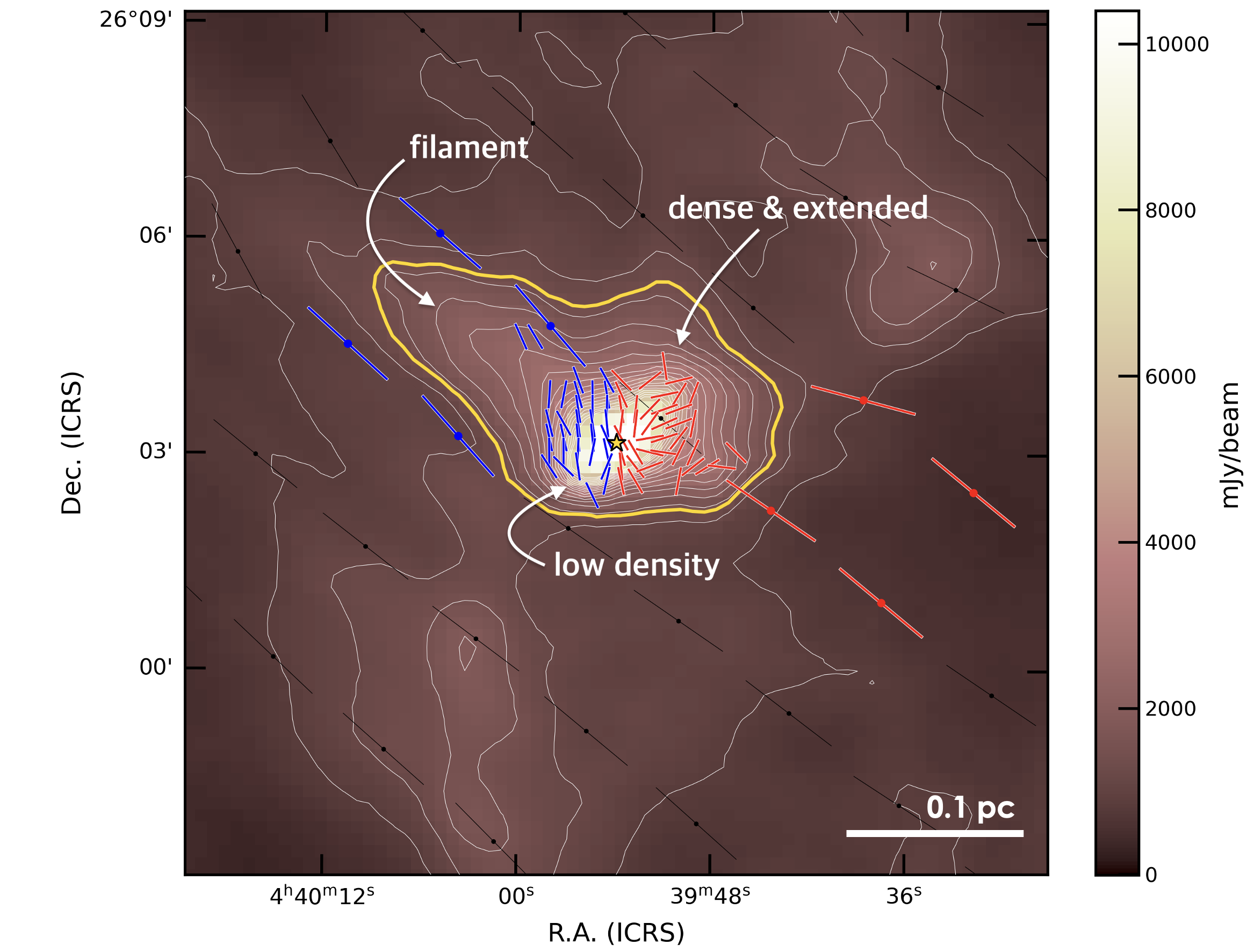}
    \caption{Herschel/SPIRE 350 $\mu$m dust continuum image overlaid with B-field orientations derived from Planck and JCMT 850 $\mu$m polarization data. Large segments with circular markers indicate the large-scale magnetic field orientations from Planck, while small segments represent the small-scale magnetic fields derived by JCMT. Blue and red segments correspond to the B-fields in the eastern and western sides, respectively. The yellow contour shows the extended cloud structure surrounding the L1527 core, and the star symbol marks the position of the protostar.}
    \label{fig:Herschel_Plank_JCMT}
\end{figure}

\begin{figure}
    \centering
    \includegraphics[width=1\linewidth]{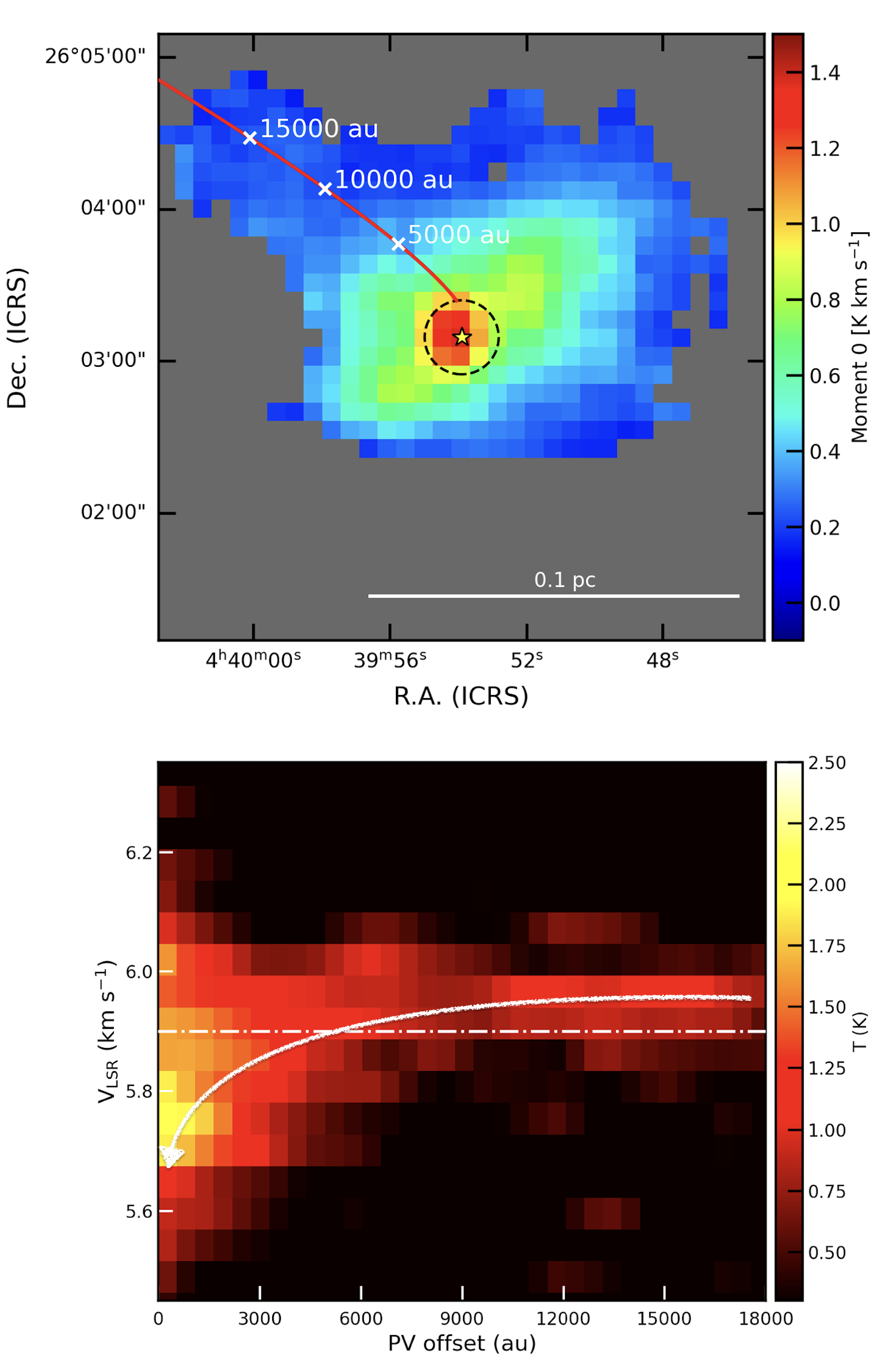}
    \caption{\textit{Top}: Integrated intensity map of C$^{18}$O ($J=3$--$2$) derived from the area of the fitted Gaussian profile at each pixel. The black dashed circle marks the central region, corresponding to twice the beam size. The red solid line shows the position--velocity (PV) cut, and the white crosses mark the projected arclength from the starting point of the line. \textit{Bottom}: Position--velocity diagram extracted along the PV cut shown in the top panel. The white dash-dotted line marks the systemic velocity of the central protostar, 5.9~km~s$^{-1}$ \citep{Tobin_2011}. White arrow presents the velocity gradient.}
    \label{fig:PV_HARP}
\end{figure}

The background intensity map in Figure~\ref{fig:Herschel_Plank_JCMT} shows the Herschel/SPIRE 350 $\mu$m dust continuum overlaid with B-field orientations derived from the Planck (large segments) and JCMT (small segments) 850 $\mu$m polarization data. Various large-scale structures are seen, which are not detected by JCMT. In particular, the L1527 protostar appears to be located on the southeastern edge of the parental cloud outlined by the yellow contour. An elongated filamentary structure extends northeastward from the L1527 core, whereas the western side shows a relatively isotropic extension.

The northeastern filamentary structure is well aligned with the surrounding large-scale magnetic fields. Near the central region, the small-scale magnetic field appears slightly distorted from the large-scale field and is oriented primarily in the north--south direction. In contrast, the large-scale magnetic field in the western region appears slightly bent toward the central core, and the small-scale field exhibits a pinched morphology aligned with the bipolar outflow.

\begin{figure*}
    \centering
    \includegraphics[width=1\linewidth]{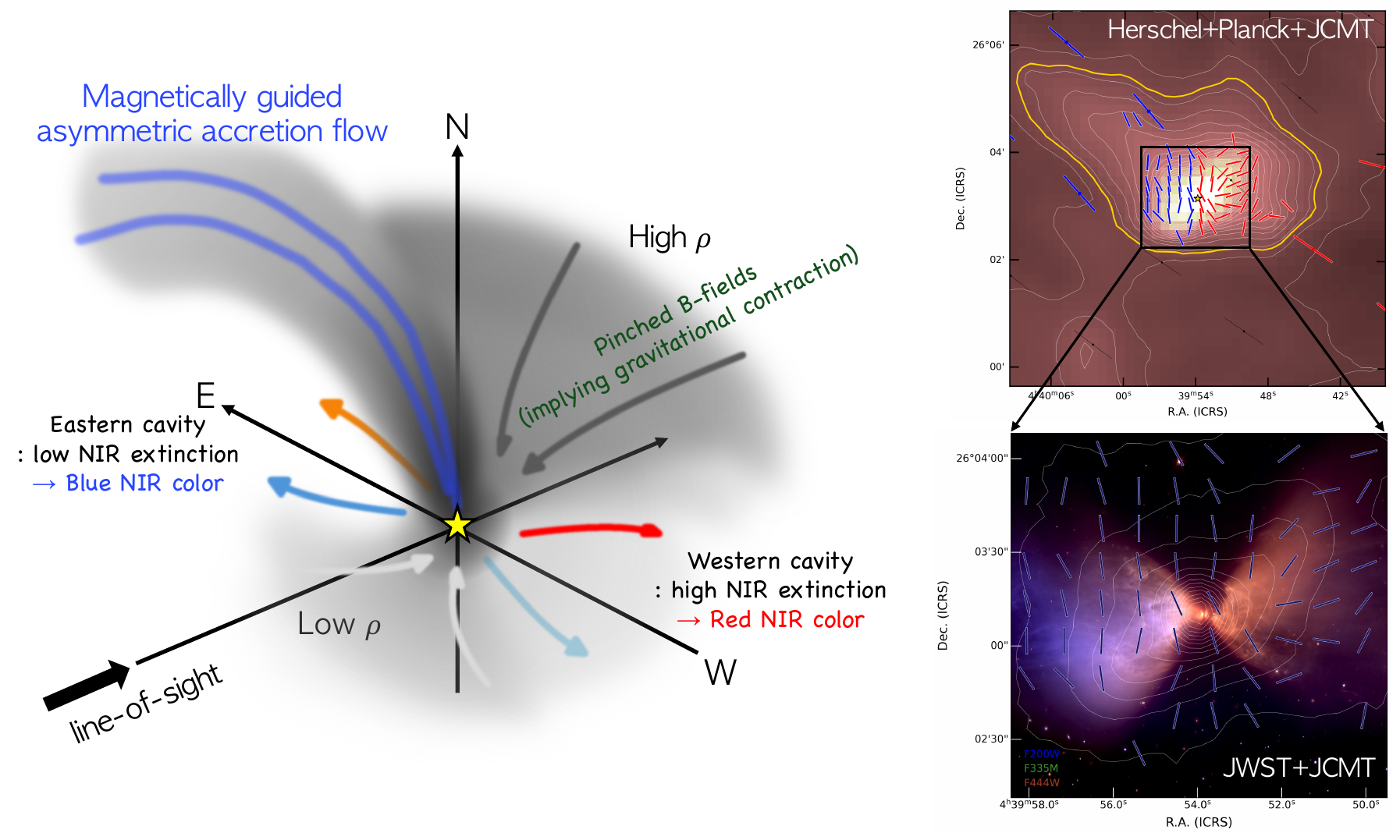}
    \caption{\textit{Left}: Schematic illustration of the mass accretion scenario in L1527. \textit{Top right}: Herschel/SPIRE 350 $\mu$m dust continuum map with magnetic field orientations from the Planck and JCMT dust polarization data. \textit{Bottom right}: RGB-composited JWST image overlaid with the JCMT magnetic fields. White contours indicate the same JCMT 850~$\mu$m intensity levels as in Figure~\ref{fig:B-vectors}.}
    \label{fig:Scenario}
\end{figure*}

We further investigated the kinematics of the northeastern filamentary structure using JCMT HARP C$^{18}$O ($J=3$--$2$) line data (Fig.~\ref{fig:PV_HARP}). The top panel of Figure~\ref{fig:PV_HARP} presents the integrated intensity map (moment 0) derived from the area of the fitted Gaussian profile at each pixel. Its spatial distribution resembles that of the dust continuum shown in Figure~\ref{fig:Herschel_Plank_JCMT}, including the northeastern filament. We extracted a position--velocity (PV) diagram along this filament (bottom panel of Fig.~\ref{fig:PV_HARP}), excluding the central region within a radius of 2000 au (marked by the black dashed circle), where multiple velocity components associated with the disk and bipolar outflow are expected to be blended. The PV diagram exhibits a smooth velocity gradient from the outer to the inner region, ranging from 5.95 km~s$^{-1}$ at the filament edge to 5.7 km~s$^{-1}$ near the center. At large distances, the gas moves inward at an approximately constant velocity, consistent with weak gravitational acceleration in the outer region. Closer to the center, the emission becomes increasingly blueshifted and eventually exceeds the systemic velocity of 5.9 km~s$^{-1}$ \citep{Tobin_2011}. Since the northern side of the L1527 protostellar disk shows redshifted rotation, this blueshifted feature in Figure~\ref{fig:PV_HARP} is unlikely to originate from rotation of the disk or inner envelope \citep[][]{Aso_2017,eDisk_L1527}. In addition, if the PV diagram were contaminated by outflow emission and the blueshifted component originated from the outflow, a corresponding redshifted component would also be expected, because the L1527 outflow lies nearly on the plane of the sky \citep[e.g.,][]{Ohashi_1997,Hogerheijde,eDisk_L1527}. However, Figure~\ref{fig:PV_HARP} shows no such redshifted counterpart, implying that the detected signal does not originate from the outflow. This type of envelope kinematic signature in the PV diagram has also been identified in various mass accretion streamers \citep[e.g.,][]{Valdivia_Mena_streamer, Kido_streamer, M512_streamer}, supporting the interpretation that the northeastern filamentary structure may be a candidate for a large-scale mass infalling flow.\\

\subsection{Star formation scenario of L1527}\label{subsec:SF_scenario}
In the previous sections, we described the large- and small-scale dust continuum structures in L1527, together with their associated magnetic fields and gas kinematics. Here, we propose one possible star formation scenario for L1527 that accounts for the overall observational results (Fig.~\ref{fig:Scenario}).

L1527 may have initially formed near the southeastern edge of the parental cloud (yellow contour in Fig.~\ref{fig:Herschel_Plank_JCMT}), where the envelope mass reservoir toward $\text{E}_{\text{out}}$ and $\text{S}_{\text{env}}$ was relatively limited, resulting in lower envelope mass and higher temperatures in these regions as inferred from the spectral index. In the eastern side, Mass accretion has primarily occured through the northeastern filamentary structure, guided by magnetic fields aligned with the filament \citep[e.g.,][]{Pillai,Kwon_2022}. Because the infall proceeds parallel to the magnetic field, deformation of the field lines remains small until the flow reaches the central region. Closer to the center, gravity and rotation become dominant and redirect the accretion flow. Since the mass accretion proceeds through a single coherent channel, the resulting deformation of the magnetic fields is relatively uniform, reorienting the magnetic fields into a nearly north--south configuration. Consequently, $\text{E}_{\text{out}}$ shows a uniform north--south magnetic field orientation. The initial lower envelope mass content in $\text{E}_{\text{out}}$ would also reduce NIR extinction, particularly at shorter wavelengths (e.g., JWST NIRCam 2~$\mu$m image in Fig.~\ref{fig:jwst_quantify}), naturally producing a relatively bluer NIR color in the eastern outflow cavity.

In contrast, \myterm{$\text{W}_{\text{out}}$ and $\text{N}_{\text{env}}$ are denser and more extended, and exhibit an hourglass-like magnetic field morphology. This pinched morphology suggests that the magnetic field structure in these regions may have been shaped by the combined effects of gravitational contraction and possible outflow-related distortion.} Consistent with this picture, $\text{W}_{\text{out}}$ and $\text{N}_{\text{env}}$ show higher density and lower temperature, as inferred from the spectral index, and appear redder in the NIR image because of the higher extinction by the dense envelope.

In terms of magnetic properties, \myterm{$\text{E}_{\text{out}}$, which is located near the connection to the larger-scale northeastern filament, has a mass-to-flux ratio of $\lambda \approx 1$ and shows a clearly sub-Alfv\'enic state ($\mathcal{M}_{A}\approx0.4$). On larger scales, the northeastern filament is well aligned with the large-scale magnetic fields and exhibits a velocity gradient in the PV diagram. Together, these properties are consistent with a scenario in which material may be guided along the asymmetric filamentary structure under the influence of the magnetic field and supplied toward the core. In contrast, $\text{W}_{\text{out}}$ and $\text{N}_{\text{env}}$ show higher mass-to-flux ratios of $\lambda \approx 2$, suggesting that gravity may play a relatively stronger role than magnetic fields in shaping the dense envelope material in these regions. Nevertheless, their clearly sub-Alfv\'enic conditions indicate that magnetic fields remain dominant over turbulence.}\\

\section{Conclusion} \label{sec:conclusion}
In this paper, we investigated the magnetic fields and physical properties of L1527 using the JCMT POL-2/SCUBA-2 instrument. The key results are summarized as follows:

\begin{enumerate}[label=(\roman*)]
    \item \textit{Magnetic fields}: L1527 shows distinct magnetic field morphologies between eastern and western outflow regions. In $\text{E}_{\text{out}}$, the magnetic fields are uniformly aligned along the north--south orientation, whereas $\text{W}_{\text{out}}$ exhibits an hourglass-like pinched morphology. Both $\text{N}_{\text{env}}$ and $\text{S}_{\text{env}}$ show radially pinched morphologies. These four regions differ in the relative importance of magnetic fields compared to gravity. \myterm{$\text{E}_{\text{out}}$ and $\text{S}_{\text{env}}$ show lower mass-to-flux ratios ($\lambda \approx 1$), whereas $\text{W}_{\text{out}}$ and $\text{N}_{\text{env}}$ show higher values ($\lambda \approx 2$). However, all regions are clearly sub-Alfv\'enic ($\mathcal{M}_{A}<0.6$).}
    \item \textit{Intensity, temperature, and column density}: L1527 shows a nearly symmetric intensity distribution between the $\text{NW}_{\text{core}}$ and southeastern $\text{SE}_{\text{core}}$, although $\text{NW}_{\text{core}}$ is more extended. The spectral index between 450 $\mu$m and 850 $\mu$m differs significantly between the regions, suggesting that $\text{E}_{\text{out}}$ and $\text{S}_{\text{env}}$ have a higher temperature and a lower density compared to $\text{W}_{\text{out}}$ and $\text{N}_{\text{env}}$.
    \item \textit{Large-scale structures}: In the Herschel/SPIRE 350 $\mu$m intensity map, the protostar appears to be located on the southeastern edge of the parental cloud. We identify a filament extending toward the northeastern direction, whereas the western side shows a more isotropic extension. The northeastern filament is well aligned with the large-scale magnetic fields and exhibits kinematic signatures of mass funneling in the C$^{18}$O observations.
    \item \textit{Star formation scenario}: In the eastern side, mass accretion occurs along the northwestern filamentary structure, guided by magnetic fields aligned with the filament. \myterm{In contrast, the dense and relatively isotropic mass distribution on the western side, together with the hourglass-like pinched magnetic field morphology, may suggest a possible role of gravitational contraction in shaping the envelope structure and magnetic fields. These anisotropic mass distribution and related accretion conditions may be associated with the distinct magnetic field patterns, as well as the contrasting temperature, density, and NIR color.} 
\end{enumerate}

\begin{acknowledgments}
We are grateful to the anonymous referee for helpful comments. This research was supported by the `Science Education in Infosphere(SEI)' funded by the four-stage BK21. W.K. is supported by the National Research Foundation of Korea (NRF) grant funded by the Korea government (MSIT) (RS-2024-00342488 and RS-2024-00416859). L.W.L. acknowledges support from NSF AST-2307844. The James Clerk Maxwell Telescope is operated by the East Asian Observatory on behalf of The National Astronomical Observatory of Japan; Academia Sinica Institute of Astronomy and Astrophysics; the Korea Astronomy and Space Science Institute; the National Astronomical Research Institute of Thailand; Center for Astronomical Mega-Science (as well as the National Key R\&D Program of China with No. 2017YFA0402700). Additional funding support is provided by the Science and Technology Facilities Council of the United Kingdom and participating universities and organizations in the United Kingdom and Canada. Additional funds for the construction of SCUBA-2 were provided by the Canada Foundation for Innovation. This research has made use of data from the Herschel Gould Belt survey (HGBS) project (http://gouldbelt-herschel.cea.fr). The HGBS is a Herschel Key Programme jointly carried out by SPIRE Specialist Astronomy Group 3 (SAG 3), scientists of several institutes in the PACS Consortium (CEA Saclay, INAF-IFSI Rome and INAF-Arcetri, KU Leuven, MPIA Heidelberg), and scientists of the Herschel Science Center (HSC). This research has made use of the NASA/IPAC Infrared Science Archive, which is funded by the National Aeronautics and Space Administration and operated by the California Institute of Technology. This work is based on observations made with the NASA/ESA/CSA James Webb Space Telescope. The data were obtained from the Mikulski Archive for Space Telescopes at the Space Telescope Science Institute, which is operated by the Association of Universities for Research in Astronomy, Inc., under NASA contract NAS 5-03127 for JWST. These observations are associated with program \#2739. All {\it JWST} data used in this paper can be found in MAST: \dataset[10.17909/yx13-xh78]{http://dx.doi.org/10.17909/yx13-xh78}.

\end{acknowledgments}

\facility{Herschel, IRSA, JCMT, JWST, Planck}
\software{Starlink \citep{Starlink}, SMURF \citep{SMURF}, NumPy \citep{numpy}, Matplotlib \citep{matplotlib}, Astropy \citep{Astropy}, SciPy \citep{Scipy}}

\appendix
\renewcommand{\thefigure}{A\arabic{figure}}
\renewcommand{\theHfigure}{A\arabic{figure}}
\setcounter{figure}{0}

\section{Comparison with previous SCUBA/SCUPOL observations}\label{app:comparison}

\begin{figure*}
    \centering
    \includegraphics[width=1\linewidth]{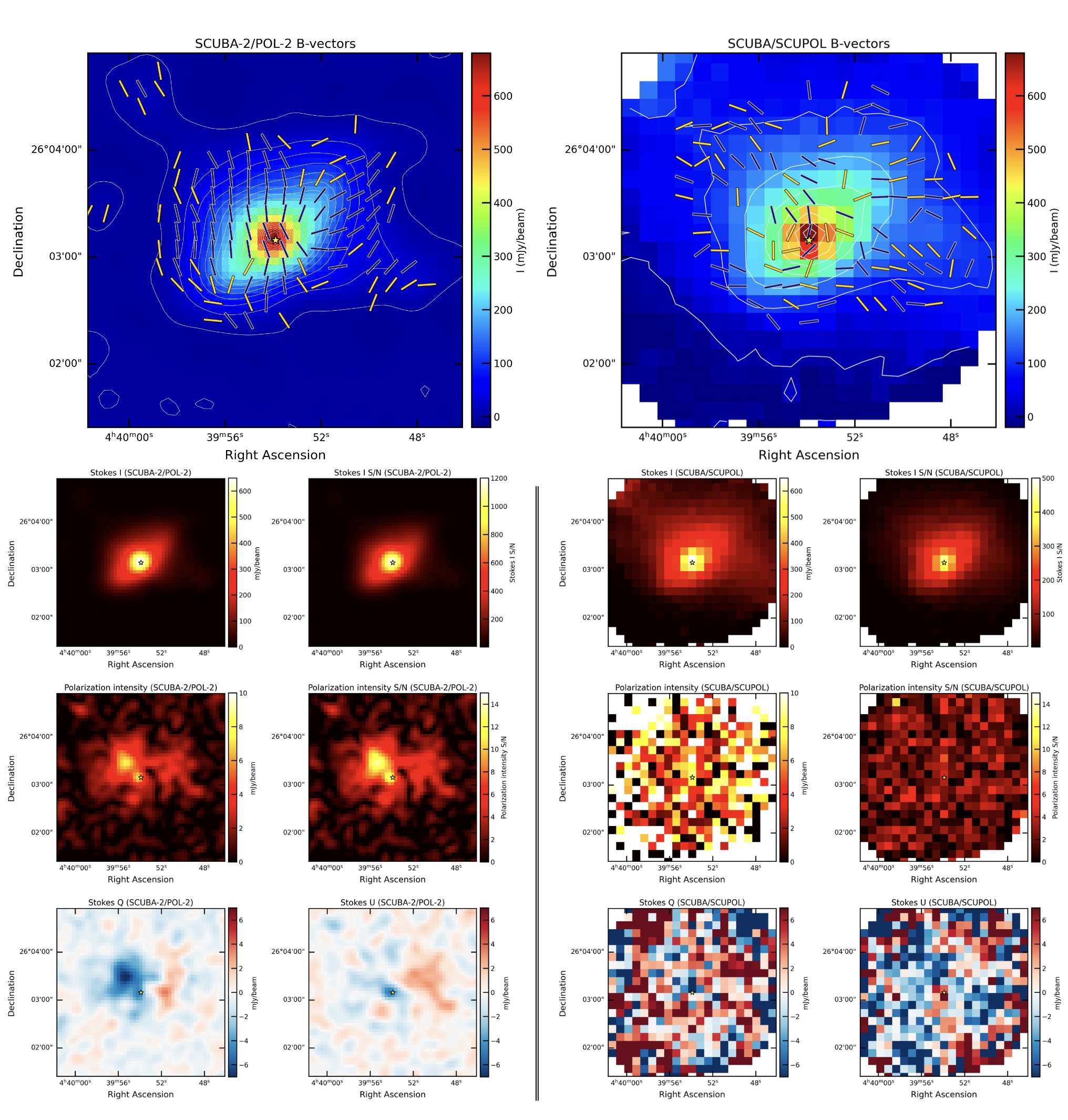}
    \caption{Comparison between the SCUBA-2/POL-2 observations and the SCUBA/SCUPOL results \citep{Matthews_SCUPOL}. \textit{Top}: Comparison of the inferred B-field orientations. Black segments indicate those with $I/\sigma_I > 3$ and $PI/\sigma_{PI} > 3$, while yellow segments indicate those with $I/\sigma_I > 3$ and $2 < PI/\sigma_{PI} \leq 3$. \textit{Bottom}: Comparison of the polarization properties, including the Stokes $I$ and polarization intensity maps with their S/N maps, as well as the Stokes $Q$ and $U$ maps.}
    \label{fig:comparison}
\end{figure*}

We compared the SCUBA-2/POL-2 observations with the previous SCUBA/SCUPOL observations to assess the improvement in the quality of the polarization data of L1527. We obtained the SCUBA/SCUPOL 850 $\mu$m data from \citet{Matthews_SCUPOL}, which combined SCUBA/SCUPOL observations taken between 1999 September 1 and 2003 December 15. The data were convolved to obtain an effective beam size of 20$^{\prime\prime}$ to achieve a higher S/N and were then sampled with 10$^{\prime\prime}$ pixels corresponding to Nyquist sampling. 

For comparison, we also smoothed our SCUBA-2/POL-2 data using a Gaussian kernel to obtain the same resolution and the corresponding noise images. We then sampled the data with 10$^{\prime\prime}$ and derived the B-field orientations. Figure~\ref{fig:comparison} presents a comparison of various polarization properties, including the B-field orientations, Stokes $I$ and polarization intensity with their S/N maps, and the Stokes $Q$ and $U$ maps.

The SCUBA/SCUPOL observations show highly random and noisy patterns in the polarization intensity, Stokes $Q$, and Stokes $U$ maps, making it difficult to identify any coherent polarization structure. Although some pixels formally satisfy the S/N threshold of polarization intensity greater than 3 (top-right panel of Fig.~\ref{fig:comparison}), they appear to originate from noisy polarization intensity maps where stochastic noise causes individual pixels to exceed the threshold.

In comparison, the SCUBA-2/POL-2 observations show clear patterns in the Stokes $Q$ and $U$ maps as well as in the polarization intensity map. In addition, the S/N of Stokes $I$ and polarization intensity is significantly higher than that in the previous observations. These results clearly demonstrate that the SCUBA-2/POL-2 observations in this study provide much more reliable measurements than the earlier SCUBA/SCUPOL observations. \\

\newpage

\section{Testing possible variations in dust opacity spectral index, $\beta$}\label{app:beta_var}
\begin{figure*}[h]
    \centering
    \includegraphics[width=0.9\linewidth]{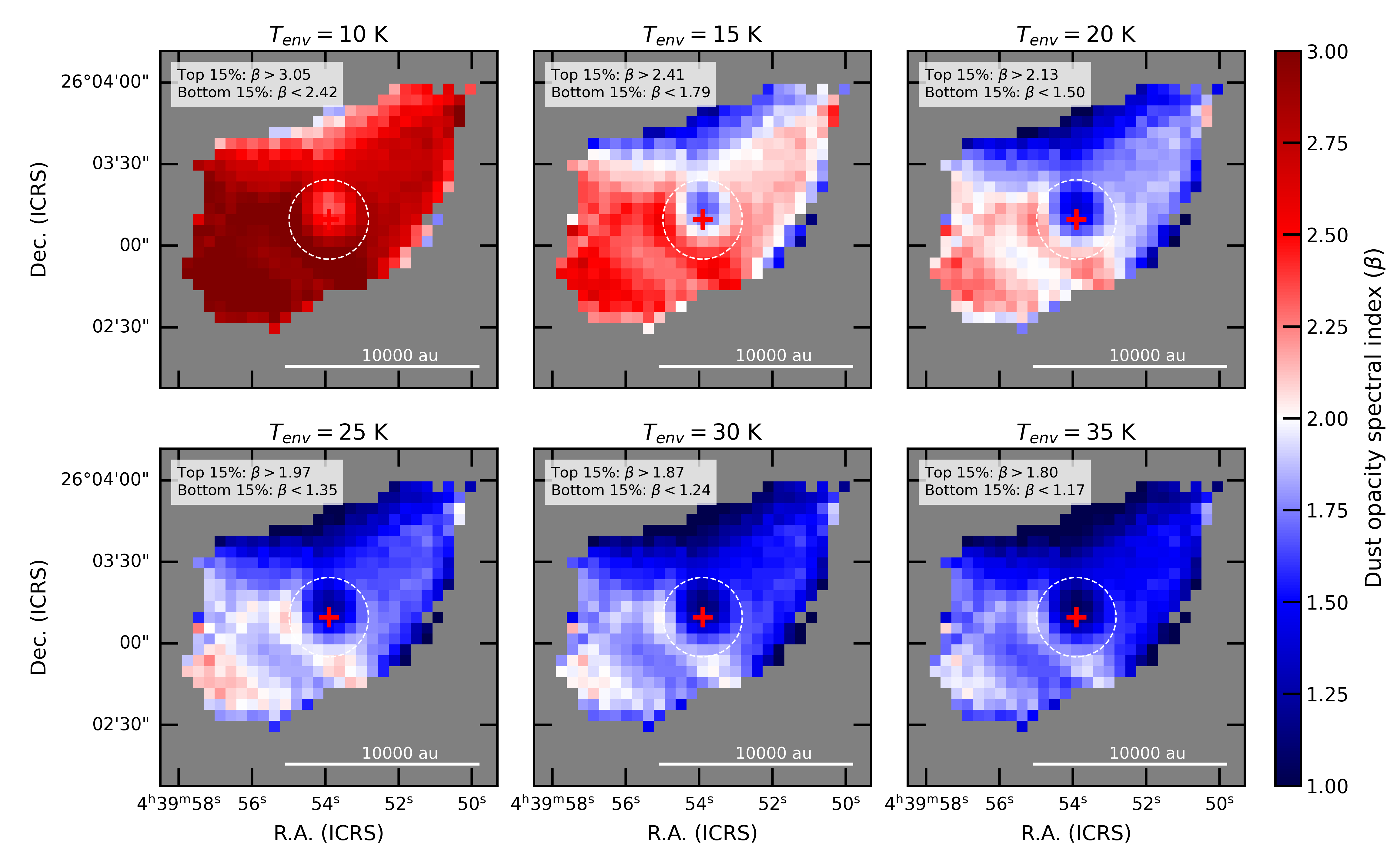}
    \caption{Maps of the dust opacity spectral index, $\beta$, derived assuming fixed temperatures from 10~K to 35~K. The highest 15\% and lowest 15\% of the $\beta$ values are indicated in each panel.}
    \label{fig:beta_var}
\end{figure*}

\myterm{To test whether the asymmetric distribution of the spectral index can be explained by variations in $\beta$ within L1527, we examined the required variation in $\beta$ over a temperature range of 10--35~K, in steps of 5~K (Fig.~\ref{fig:beta_var}). We found that a variation in $\beta$ of at least 0.6 is required to reproduce the observed spectral index asymmetry. Moreover, for all temperatures considered, some regions require either significantly low ($\beta<1.4$) or high ($\beta>2.4$) values. Therefore, we conclude that the observed spectral-index asymmetry is more likely to arise from an asymmetric temperature distribution than from intrinsic variations in dust properties.}\\


\bibliography{L1527}{}
\bibliographystyle{aasjournal}



\end{document}